\documentclass[a4paper,12pt]{article}
\pdfoutput=1
\usepackage{feynmp-auto,expdlist}
\usepackage{amsmath, amsfonts, amssymb}
\usepackage{graphicx}
\usepackage{enumerate}
\usepackage{hyperref}
\usepackage{latexsym}
\usepackage{dsfont}
\usepackage{hepnicenames}
\usepackage{enumerate}
\usepackage{soul}
\usepackage[normalem]{ulem}
\usepackage{bbold}

\usepackage{units}

\newcommand{\Q}{{\cal Q}}

\usepackage{mathrsfs,graphicx,rotating,amsmath,amsfonts,mathtools,booktabs,amssymb,wasysym}
\usepackage{hyperref}\usepackage{slashed}
\usepackage[nosort]{cite}
\usepackage[table,xcdraw,dvipsnames]{xcolor}
\usepackage{bm}
\usepackage{graphicx}
\usepackage{multirow,multicol}
\hypersetup{colorlinks,bookmarksopen,bookmarksnumbered,
	linkcolor=blus,pdfstartview=FitH,urlcolor=rossos,citecolor=verde}
\allowdisplaybreaks

\newcommand{\GammaDG}{\Gamma_{\scriptscriptstyle{\rm DG}}}
\renewcommand{\[}{\left[}

\renewcommand{\(}{\left(}
\renewcommand{\)}{\right)}
\renewcommand{\S}{{\cal S}}

\def\Lag{\mathscr{L}}

\newcommand{\mio}[1]{}

\newcommand{\med}[1]{\langle #1\rangle}

\def\bpm{\begin{pmatrix}}
	\def\epm{\end{pmatrix}}

\usepackage{mathrsfs}

\newcommand{\fig}[1]{~\ref{fig:#1}}
\newcommand{\sfrac}[2]{#1/#2}

\allowdisplaybreaks
\usepackage{multicol}
\usepackage{color}
\definecolor{rosso}{cmyk}{0,1,1,0.4}
\definecolor{rossos}{cmyk}{0,1,1,0.55}
\definecolor{rossoc}{cmyk}{0,1,1,0.2}
\definecolor{blu}{cmyk}{1,1,0,0.3}
\definecolor{blus}{cmyk}{1,1,0,0.6}
\definecolor{bluc}{cmyk}{1,1,0,0.1}
\definecolor{verde}{cmyk}{0.92,0,0.59,0.25}
\definecolor{verdec}{cmyk}{0.92,0,0.59,0.15}
\definecolor{verdes}{cmyk}{0.92,0,0.59,0.4}

\newcommand{\eq}[1]{~{\rm (\ref{eq:#1})}}

\newcommand{\GeV}{\,{\rm GeV}}
\newcommand{\TeV}{\,{\rm TeV}}

\newcommand{\Tr}{\,{\rm Tr}}
\newcommand{\diag}{\,{\rm diag}}

\def\circa#1{\,\raise.3ex\hbox{$#1$\kern-.75em\lower1ex\hbox{$\sim$}}\,}

\newcommand{\beq}{\begin{equation}}
\newcommand{\eeq}{\end{equation}}

\newcommand{\bea}{\begin{eqnarray}}
\newcommand{\eea}{\end{eqnarray}}
\newcommand{\be}{\begin{equation}}
\newcommand{\ee}{\end{equation}}
\font\tenrsfs=rsfs10 at 12pt
\font\sevenrsfs=rsfs7
\font\fiversfs=rsfs5
\newfam\rsfsfam
\textfont\rsfsfam=\tenrsfs
\scriptfont\rsfsfam=\sevenrsfs
\scriptscriptfont\rsfsfam=\fiversfs

\newcommand{\D}{{\cal D}}

\newsavebox\MBox

\newcommand{\Sp}{\,{\rm Sp}}

\newcommand{\SU}{\,{\rm SU}}
\newcommand{\SO}{\,{\rm SO}}
\newcommand{\U}{\,{\rm U}}
\renewcommand{\L}{\mathscr{L}}

\def\circa#1{\,\raise.3ex\hbox{$#1$\kern-.75em\lower1ex\hbox{$\sim$}}\,}
\makeatletter

\font\ital=cmu10

\def\hhref#1{\href{http://arxiv.org/abs/#1}{arXiv:#1}}
\usepackage{xstring}
\newcommand{\hhrefq}[1]{\IfSubStr{#1}{:}{\href{http://inspirehep.net/search?ln=en&ln=en&p=#1&of=hb&action_search=Search&sf=&so=d&rm=&rg=25&sc=0}{InSpire:#1}}{\hhref{#1}}}

\def\art{\@ifnextchar[{\eart}{\oart}}
\def\eart[#1]#2#3#4#5#6{{\rm #2}, {\em #3 \bf #4} {\rm (#6) #5} ({\em #1})}
\def\article{\@ifnextchar[{\earticle}{\oarticle}}
\def\oarticle#1#2#3#4#5#6{{\rm #1}, {\ital ``#6''}, {\rm #2 #3 (#5) #4}}
\def\earticle[#1]#2#3#4#5#6#7{{\rm #2}, {\ital ``#7''}, {\rm #3 #4 (#6) #5}  [\hhrefq{#1}]}
\def\hepart[#1]#2{{\rm #2, \sl#1}}
\def\heparticle[#1]#2#3{#2, {\ital ``#3''} [\hhrefq{#1}]}
\newcommand{\doi}[1]{\href{http://dx.doi.org/#1}{[link]}}

\newcommand{\hhrefqq}[1]{\IfBeginWith{#1}{10.}{\href{https://doi.org/#1}{doi:#1}}{\hhrefq{#1}}}
\def\earticle[#1]#2#3#4#5#6#7{{\rm #2}, {\ital ``#7''}, {\rm #3 #4 (#6) #5}  [\hhrefqq{#1}]}

\newcounter{alphaequation}[equation]
\def\thealphaequation{\theequation\hbox to
	0.6em{\hfil\alph{alphaequation}\hfil}}
\def\eqnsystem#1{
	\def\@eqnnum{{\rm (\thealphaequation)}}
	\def\@@eqncr{\let\@tempa\relax \ifcase\@eqcnt \def\@tempa{& & &} \or
		\def\@tempa{& &}\or \def\@tempa{&}\fi\@tempa
		\if@eqnsw\@eqnnum\refstepcounter{alphaequation}\fi
		\global\@eqnswtrue\global\@eqcnt=0\cr}
	\refstepcounter{equation} \let\@currentlabel\theequation \def\@tempb{#1}
	\ifx\@tempb\empty\else\label{#1}\fi
	\refstepcounter{alphaequation}
	\let\@currentlabel\thealphaequation
	\global\@eqnswtrue\global\@eqcnt=0 \tabskip\@centering\let\\=\@eqncr
	$$\halign to \displaywidth\bgroup \@eqnsel\hskip\@centering
	$\displaystyle\tabskip\z@{##}$&\global\@eqcnt\@ne
	\hskip2\arraycolsep\hfil${##}$\hfil& \global\@eqcnt\tw@\hskip2\arraycolsep
	$\displaystyle\tabskip\z@{##}$\hfil
	\tabskip\@centering&\llap{##}\tabskip\z@\cr}
\def\endeqnsystem{\@@eqncr\egroup$$\global\@ignoretrue} \makeatother

\definecolor{Gray}{gray}{0.95}

\def\bal#1\eal{\begin{align}#1\end{align}}

\newcommand{\W}{\mathcal{W}}
\newcommand{\Z}{\mathcal{Z}}

\newcommand{\A}{\mathcal{A}}

\newcommand{\N}{\mathcal{N}}

\newcommand{\g}{\mathfrak{g}}
\newcommand{\M}{\mathcal{M}}
\newcommand{\G}{\mathcal{G}}

\renewcommand{\L}{{\cal L}}
\renewcommand{\P}{{\cal P}}
\newcommand{\MLL}{M_{\L\L}}

\begin{document}
\vspace{1.5cm}

\begin{center}
{\Large\LARGE\Huge \bf \color{rossos} 
Axion quality from the (anti)symmetric of $\SU(\N)$
}\\[1cm]
{\bf Marco Ardu$^{a}$, Luca Di Luzio$^{b}$, Giacomo Landini$^{a,c}$, \\
Alessandro Strumia$^{a}$, Daniele Teresi$^{a,c}$, Jin-Wei Wang$^{a,d,e}$}\\[7mm]

{\it $^a$ Dipartimento di Fisica, Universit\`a di Pisa, Italy}\\[1mm]
{\it $^b$ DESY, Notkestra\ss e 85, D-22607 Hamburg, Germany}\\[1mm]
{\it $^c$ INFN, Sezione di Pisa, Italy}\\[1mm]
{\it $^d$ Key Laboratory of Particle Astrophysics, Institute of High Energy Physics,
	Chinese Academy of Sciences, Beijing, China}\\[1mm]
{\it $^e$ School of Physical Sciences, University of Chinese Academy of Sciences, Beijing, China}\\[1mm]

\vspace{0.5cm}

\begin{quote}\large
We propose two models where a U(1) Peccei-Quinn 
global symmetry arises accidentally and is respected up to high-dimensional
operators, so that the axion 
solution to the strong CP problem 
is successful even in the presence of Planck-suppressed operators.
One model is $\SU(\N)$ gauge interactions with fermions in the fundamental 
and a scalar in the symmetric. The axion arises from
spontaneous symmetry breaking to $\SO(\N)$, that confines at a lower energy scale.
Axion quality in the model needs $\N\circa{>}10$. SO bound states and possibly monopoles provide
extra Dark Matter candidates beyond  the axion. 
In the second model the scalar is in the anti-symmetric:
$\SU(\N)$ broken to $\Sp(\N)$ needs even $\N\circa{>}20$. 
The cosmological DM abundance, consisting of axions and/or super-heavy relics, can be reproduced if the PQ symmetry is broken before inflation
(Boltzmann-suppressed production of super-heavy relics)
or after (super-heavy relics in thermal equilibrium get partially diluted by dark glue-ball decays).


\end{quote}

\end{center}

\setcounter{footnote}{0}

\newpage

\tableofcontents

\section{Introduction}


The Peccei-Quinn (PQ) solution to the strong CP problem~\cite{Peccei:1977hh,Peccei:1977ur} 
has a problematic aspect: it relies on a global $\U(1)_{\rm PQ}$ symmetry 
which, although broken at low energy by the QCD anomaly, 
must be an extremely good symmetry of high-energy
physics. 
This issue is known as the \emph{PQ quality problem} 
\cite{Georgi:1981pu,Dine:1986bg,Barr:1992qq,Kamionkowski:1992mf,Holman:1992us,Ghigna:1992iv}. 
Global symmetries are believed not to be fundamental, 
and arise as accidental symmetries e.g.\ in gauge theories. 
Well known examples are baryon and lepton numbers in the Standard Model (SM). 
Conceptually, there are two steps in the formulation of the problem:  
\begin{itemize}
\item[$i)$] the $\U(1)_{\rm PQ}$ should arise accidentally in a renormalizable Lagrangian; 
\item[$ii)$] approximating higher-energy physics as non-renormalizable operators suppressed by 
some scale $\Lambda_{\rm UV}$, the $\U(1)_{\rm PQ}$ should be preserved 
by operators with dimension up to $d \sim 9$ 
assuming $\Lambda_{\rm UV} \sim M_{\rm Pl}$ and an 
axion decay constant $f_a \gtrsim 10^9$ GeV. 
\end{itemize}
The bound becomes stronger for higher $f_a$ and lower $\Lambda_{\rm UV}$. 
Indeed, it comes from requiring that the energy density 
due to UV sources of $\U(1)_{\rm PQ}$ breaking is 
about 
$10^{-10}$ 
times smaller than the energy density 
of the QCD axion potential
\beq 
\label{eq:PQestimate}
\( \frac{f_a}{\Lambda_{\rm UV}} \)^{d-4} f_a^4 \lesssim 10^{-10} \Lambda^4_{\rm QCD} ,
\eeq
so that the induced axion vacuum expectation value (VEV) 
is $\med{a} / f_a \lesssim 10^{-10}$, within the neutron 
electric dipole moment bound. 

In string models one expects towers of new states below or around the Planck scale, 
potentially generating PQ-breaking higher-dimensional operators,
that make manifest the PQ-quality problem.
Furthermore, it is believed that gravity violates global symmetries, 
based on semi-classical arguments related to black holes and Hawking radiation. 
In scenarios in which Einstein gravity is minimally
coupled to the axion field, non-conservation of the PQ global charge arises 
from non-perturbative effects described by Euclidean wormholes.
These effects are calculable to some extent
and correct the axion potential as~\cite{Abbott:1989jw,Coleman:1989zu,hep-th/9502069,Alonso:2017avz,1807.00824,2009.03917}
\beq 
\label{eq:whpotential}
\sim M_{\rm Pl}^4 e^{-S_{\rm wh}} \cos\(a+\delta\) ,
\eeq
where $\delta\sim 1$ is a generic displacement due to the fact that the gravity contribution 
does not need to be aligned to the low-energy QCD contribution. 
Computing the wormhole action taking into account the axion only gives
$S_{\rm wh} \sim N M_{\rm Pl} / f_a$, so that the contribution in eq.~\eqref{eq:whpotential}
poses a problem for the PQ solution if $f_a/N \gtrsim 6~ 10^{16}$ GeV,
where the integer $N$ is the minimal PQ charge carried by the wormhole.
In theories where $f_a$ is the vacuum expectation value of some sub-Planckian field, 
this grows reaching the Planck scale in the wormhole throat,
giving a reduced $S_{\rm wh} \sim  N \ln M_{\rm Pl} / f_a$~\cite{hep-th/9502069}.
According to~\cite{Abbott:1989jw,hep-th/9502069,2009.03917} this is equivalent 
to local operators with an extra suppression
$e^{-S_{\rm wh}}  \sim (f_a/M_{\rm Pl})^N$
with respect to the generic Planck-suppressed operators considered in this paper.


Eq.\eq{whpotential} holds if gravity is well described by the Einstein term at Planckian energies.
An alternative possibility is that gravity gets modified at lower energies where it is still 
weakly coupled
so that it remains weakly coupled, making non-perturbative effects irrelevant.
This for example arises in 4-derivative gravity, a renormalizable theory that allows for
accidental global symmetries not broken by higher dimensional operators and negligibly broken by
non-perturbative gravitational effects~\cite{Salvio:2014soa}. Such theory, however, contains potentially problematic
negative kinetic energy at the classical level (see e.g.~\cite{Gross:2020tph}).

We here assume that the PQ-quality problem is a real problem  and address it 
by devising a simple gauge dynamics 
along the lines of \cite{DiLuzio:2017tjx,Buttazzo:2019mvl}
that gives an accidental global PQ symmetry respected
by operators up to large enough dimension. 
Different approaches to the PQ quality problem, 
but also based on non-abelian gauge dynamics, 
have been discussed in~\cite{Randall:1992ut,Dobrescu:1996jp,1602.05427,1811.03089,1811.04039,1812.08174}.  
Section~\ref{Sp} describes a model based on a gauge group $\SU(\N)$ spontaneously broken to $\Sp(\N)$
by a scalar $\S$ in the anti-symmetric representation in the presence of fermions in the fundamental, as listed in table~\ref{tab:reps}. 
The PQ symmetry is broken by the $\N/2$-dimensional local operator $\sqrt{\det\S}$.
Section~\ref{SO} considers a similar model where a scalar 
$\S$ in the symmetric breaks 
$\SU(\N)\to \SO(\N)$, and the first PQ-breaking operator is 
the $\N$-dimensional operator $\det\S$.
Both models can provide extra Dark Matter (DM) candidates beyond the axion.
Section~\ref{model} outlines some common features of the two models.
Section~\ref{Sp} describes the model with a scalar in the anti-symmetric,
and section~\ref{SO} the model with a scalar in the symmetric.
Conclusions are given in section~\ref{concl}.

\section{Outline of the models}\label{model}
\begin{table}[t]
$$\begin{array}{c|c|cccc|ccc}
\rowcolor[HTML]{C0C0C0} 
\hbox{Field}&\hbox{Lorentz} & \multicolumn{4}{|c|}{\hbox{Gauge symmetries}} &
 \multicolumn{3}{|c}{\hbox{Global accidental symmetries}}\\
\rowcolor[HTML]{C0C0C0} 
\hbox{name} & \hbox{spin} &\U(1)_Y &  \SU(2)_L & \SU(3)_c &  \SU(\N)  & \U(1)_{\rm PQ} &  \U(1)_{\Q}&  \U(1)_{\L} \\ \hline
\S  &0&0&1&1& \N \N& +1& 0& 0 \\  \hline
\Q_L &1/2 &+Y_\Q&1& 3 & \N  & +1/2& +1& 0 \\ 
\Q_R &1/2 &-Y_\Q&1& \bar 3 &  \N & +1/2& -1& 0 \\
\L^{1,2,3}_L &1/2&+Y_\L&1& 1 & \bar \N  & -1/2& 0& +1 \\
\L^{1,2,3}_R &1/2&-Y_\L&1& 1& \bar \N  & -1/2& 0& -1 \\ 
\end{array}$$
\caption{\em Field content of the model.
The scalar $\S$ can be in the anti-symmetric (section~\ref{Sp}) or in the symmetric (section~\ref{SO})
two-index representation.
The heavy quarks $\Q_L,\Q_R$ and leptons $\L_L,\L_R$ are Weyl doublets.
If $\L$ have vanishing hypercharge, their bound states could become acceptable DM candidates and
there is no difference between $\L_L$ and $\L_R$.
\label{tab:reps}}
\end{table}%

We consider a gauge group $G_{\rm SM}\otimes\SU(\N)$,
with a new scalar $\S$ in the two-index
symmetric or anti-symmetric representation of $\SU(\N)$,
and new left-handed chiral Weyl fermions charged under $\SU(\N)$
as listed in table~\ref{tab:reps}: 
one $\Q$ dubbed `quark'  because in the fundamental of color,
and three $\L$ dubbed `leptons'  because uncolored.\footnote{The fermion content shares some similarities with some of the composite accidental axion models of \cite{1602.05427}. The main qualitative difference is that there their $\L$ form a triplet under an extra $\SU(3)$ and QCD is the vectorial subgroup of the two $\SU(3)$ factors. This gives a different anomaly structure such that only in our case large $\N$ leads to PQ quality.}
Three $\L$ are needed in order to avoid gauge anomalies and to obtain the desired PQ anomalies.
The three $\L$ could be $1\oplus 2$ or $1\oplus 1\oplus 1$ under $\SU(2)_L$; 
as the choice does not make a big difference we assume the latter possibility and that
all 3 leptons have the same hypercharge $Y_\L$, for the moment left unspecified and possibly vanishing.
Irrespectively of their hypercharges, the fermions $\Q$ and $\L$ are chiral: their
mass terms are forbidden by gauge invariance for all values of
the hypercharges $Y_\Q$ and $Y_\L$.




As discussed in the next sections, the renormalizable theory contains three accidental global U(1) symmetries:
the one acting as a phase rotation of the scalar $\S$ will be the PQ symmetry.
It gets spontaneously broken by the vacuum expectation value of
the scalar $\S$, that also breaks $\SU(\N)$ to either $\Sp(\N)$ 
(scalar in the anti-symmetric, studied in section~\ref{Sp}) or to $\SO(\N)$ 
(scalar in the symmetric, studied in section~\ref{SO}).
As a result the fermions $\Q$ and $\L$ acquire mass
from Yukawa couplings to $\S$ and
the phase of $\S$ becomes the axion. 
As $\N =\bar\N$ for $\SO(\N)$ and $\Sp(\N)$, their condensation at
lower energy preserves SM gauge symmetries.
In both models all gauge anomalies vanish, and $\U(1)_{\text{PQ}}$
has the desired anomalies:
\begin{itemize}

\item There is a non-vanishing $\U(1)_{\rm PQ} \SU(3)_c^2$ anomaly:
to achieve this we introduced the 
fermions $\L$ and $\Q$ in two different representations of color.
We chose the simplest ones (singlet and triplet), although
different models using more complicated representations of color are possible.

\item The $\U(1)_{\rm PQ} \U(1)_Y^2$ anomaly is proportional to $Y_\Q^2-Y_\L^2$ and contributes to the axion-photon coupling.

\item We introduced the appropriate number of leptons $\L$ such that
the $\U(1)_{\rm PQ} \SU(\N)^2$  anomaly vanishes: then
the axion relaxes the $\SU(3)_c$ $\theta$ term, rather than the one of the extra gauge group $\SU(\N)$.
\end{itemize}
A similar 
model 
based on 
$\SU(\N)_L \times \SU(\N)_R$ gauge dynamics 
broken by a 
scalar transforming in the 
bi-fundamental 
down to $\SU(\N)_{L+R}$
was considered in \cite{DiLuzio:2017tjx} 
(see also \cite{Nardi:2011st,Espinosa:2012uu,Fong:2013dnk}), 
which shares similarities with the two models presented here.   
Differently from \cite{DiLuzio:2017tjx}, we assign non-zero SM 
hypercharges to the exotic fermions and show that it is possible to get rid of dangerous colored relics. This enlarges the parameter space of the model 
also to the case where the PQ is broken after inflation and 
opens the possibility of having 
extra DM candidates in the form of Sp/SO bound states. 

\section{Antisymmetric scalar that breaks $\SU(\N)\to\Sp(\N)$}\label{Sp}
We assume even $\N$, as for odd $\N$  symmetry breaking is slightly different and the axion
is eaten by a vector \cite{Buttazzo:2019mvl}.
If $\N> 8$
 the most generic renormalizable Lagrangian is
\begin{equation}\label{lagrangian}
\Lag=\Lag_{\rm SM}+\Lag_{\rm kin} + \Lag_{\rm Yuk}-V(\S).
\end{equation}
Using Weyl two-component spinors 
\begin{eqnsystem}{sys:Lags}\label{lagrangian2}
	\Lag_{\rm kin} &=&
	-\frac{1}{4}{\G^{A \mu\nu}}{\G_{\mu\nu}^A}+
	\Tr(\D_\mu \S)(\D^\mu \S)^\dagger + \sum_{f=\Q_{L,R},\L_{L,R}}\bar f i  D_\mu \sigma^\mu f\\
-\Lag_{\rm Yuk} &=& \left\{\begin{array}{ll}
y_\Q\,  \Q_L \S^* \Q_R + y_\L^{ij} \,\L_L^i \S \L_R^j + \text{h.c.}  & \hbox{if $Y_\L\neq 0$}\\
 y_\Q\,  \Q_L \S^* \Q_R +y^{ii'}_\L \,\L^i \S \L^{i'}/{2} + \text{h.c.} & \hbox{if $Y_\L= 0$}
 \end{array}\right.  \label{eq:YukSp}\\
V(\S)&=&M_\S^2 \Tr (\S\S^\dagger)+\lambda_\S \Tr (\S\S^\dagger)^2+\lambda_\S'\Tr(\S\S^\dagger \S\S^\dagger)-\lambda_{HS}(H^\dagger H)\Tr(\S\S^\dagger),
\end{eqnsystem}
where $H$ is the SM Higgs doublet.
If $Y_\L\neq 0$ without loss of generality we can rotate to a basis where the Yukawa matrix $y_\L$ is diagonal,
$\diag (y_{\L_1}, y_{\L_2}, y_{\L_3})$, with real positive entries.
If $Y_\L=0$ the matrix $y_\L$ is anti-symmetric and can be rotated to
$\diag (y_{\L_1}, y_{\L_2}, y_{\L_3})\otimes \epsilon$ where $\epsilon$  is the $2\times 2$
antisymmetric Levi-Civita tensor.

\subsubsection*{Accidental symmetries}
The gauge-covariant kinetic terms
are invariant under phase rotations of each field.
In the presence of the Yukawa and potential couplings the theory remains accidentally invariant under
\beq \label{u(1)simm}
\U(1)_{\Q} \otimes \U(1)_{\L_{1,2,3}} \otimes \U(1)_{\rm PQ}
\eeq 
where $\U(1)_{\Q}$ and $\U(1)_{\L_i}$ are the baryon and lepton 
numbers of $\Q$ and $\L_i$ according 
to which $\Q_L$ and $\Q_R$ have the opposite charge  (similarly for leptons), 
while $\S$ is uncharged. The $\U(1)_{\rm PQ}$ symmetry acting on $\S$ can be identified (a posteriori) 
as a PQ symmetry and it acts as shown in table~\ref{tab:reps},
where we chose a convenient basis.
The accidental flavour symmetry rotates with opposite phases the two $\L$ fields involved in each mass term:
for $Y_\L\neq 0$ mass terms involve $\L_L\L_R$ pairs, while for $Y_\L=0$
a similar pair structure arises at renormalizable level thanks to the anti-symmetry of the mass matrix.


\subsubsection*{Landau poles}\label{LP}
We constrain the field content and parameters of the model by 
requiring that its couplings
do not hit Landau poles below the Planck scale.
The $\SU(\N)$ gauge coupling $\g$ is asymptotically free.
Above the masses $m_\Q , m_\L$ of the new fermions,
the one-loop beta functions of the strong and hypercharge gauge coupling $g_1^2=5g_Y^2/3$ are
\beq
\frac{dg_3^2}{d\ln \mu^2}=\frac{g_3^4}{(4\pi)^2} \left( -7+\frac{2}{3}\N\right),\qquad
\frac{dg_1^2}{d\ln \mu^2}=\frac{g_1^4}{(4\pi)^2} \left( \frac{41}{10}+\frac{12\N}{5}(Y_\L^2+Y_Q^2) \right).\eeq
Assuming 
$m_\Q \sim m_\L \sim 10^{11}$GeV,
sub-Planckian Landau poles in $g_3$ and $g_Y$ are avoided if
$\N \lesssim 30$ and 
$\N (Y_\Q^2 + Y_\L^2)\lesssim 4$.

\subsection{Symmetry breaking and perturbative spectrum}
In a range of potential parameters, the scalar $\S$ acquires vacuum expectation value
$\med{\S}=w\gamma_{\N}$ where $\gamma_{\N}=\mathbb{1}_{\N/2}\otimes \epsilon$ is the invariant tensor under symplectic transformations.
This breaks $\SU(\N)\otimes\U(1)_\text{PQ}$ to $\Sp(\N)$ leaving one axion and giving mass to all new fermions. 
Following~\cite{Buttazzo:2019mvl} for even $\N$ the scalar field is conveniently parametrised as
\beq
\S=\left[\left(w+\frac{s}{\sqrt{\N/2}}\right)\gamma_{\N}+2(\Tilde{s}^b+i\Tilde{a}^b)\tilde{T}^b\gamma_{\N}\right]e^{\frac{ia}{\sqrt{\N/2}w}},
\eeq
where $\tilde{T}^b$ are the $\SU(\N)$ generators such that $\gamma_\N \tilde{T}^b$ is anti-symmetric (which satisfy the condition $\tilde{T}^*=-\gamma_\N\tilde{T}\gamma_\N$ and corresponds to the broken generators). 
The mass spectrum at perturbative level is:
\begin{itemize}
\item $\N(\N+1)/2$ massless vectors $\mathcal{A}^a$ in the adjoint of $\Sp(\N)$;
\item $\N(\N-1)/2-1$ vectors $\mathcal{W}^b$ in the traceless anti-symmetric of $\Sp(\N)$, that acquire a squared mass $M^2_{\W}=\g^2w^2$ eating the Goldstone bosons $\tilde{a}^b$;
\item The massless scalar $a$, singlet under $\Sp(\N)$. 
In view of the 
$\U(1)_{\rm PQ} \SU(3)_c^2$ anomaly 
it can be called axion and, as shown below, 
its decay constant will be $f_a=w/\sqrt{2\N}$.
The axion will get a mass because of the QCD anomaly;
\item The scalar $s$ singlet under $\Sp(\N)$. 
If symmetry breaking arises through the Coleman-Weinberg mechanism it is light
with squared mass $M^2_s=24(\N \lambda_\S +\lambda_\S')w^2$;
\item $\N(\N-1)/2-1$ scalars $\tilde{s}^b$ with squared mass $M^2_{\tilde{s}}=8(\N\lambda_\S+3\lambda_\S')w^2$, in the traceless anti-symmetric of $\Sp(\N)$;
\item One colored Dirac quark $\Psi_\Q =( \gamma_\N \Q_L ,\bar{\Q}_R)^T$ 
with mass $M_{\Q}=y_\Q w$ in the anti-fundamental of $\Sp(\N)$ charged under the accidental global  $\U(1)_\Q$; 
\item Three Dirac leptons with masses $M_{\L^i}=y_{\L}^i w$ in the fundamental of $\Sp(\N)$ and
charged under the accidental global $\U(1)_{\L_i}$.
Dirac fermions are constructed pairing the Weyl fermions involved in each mass term e.g.\
$\Psi_\L^i = (\gamma_\N \L_L^i ,\bar{\L}_R^i)$ if  $Y_\L\neq 0$. 

\end{itemize}
The fermions are perturbatively stable thanks to the unbroken global $\U(1)_{\Q,\L_i}$ symmetries discussed in eq.~\eqref{u(1)simm}.

\subsection{Confinement and bound states}
The $\Sp(\N)$ gauge dynamics confines at the energy scale 
\begin{equation}
\Lambda_{\Sp}=f_a\exp\left[-\frac{12\pi}{11(\N+2)\alpha_{\rm DC}(f_a)}\right],
\end{equation}
where $\alpha_{\rm DC}=\g^2/4\pi$ and we took into account only the running due to the gauge bosons $\A$. 
We normalise the Dynkin index of $\SU(\N)$ as 
$S_2 = 1/2$ for the fundamental.
In the confined phase, the baryons containing fermions 
($\epsilon_\N \mathcal{W}^{(\N-2)/2}\Q\Q$, $\epsilon_\N \mathcal{W}^{(\N-2)/2}\L\L$ and $\epsilon_\N \mathcal{W}^{(\N-2)/2}\Q\L$, 
with $\epsilon_\N$ denoting the $\N$-dimensional Levi-Civita tensor)
decay into lighter mesons $\Q\gamma_\N\Q$, $\L\gamma_\N\L$ and $\Q\L$. 
Depending on the constituent masses $m_\Q$ and $m_\L$, 2 or 3 of such mesons are stable because of the accidental
fermion-number symmetries $\U(1)_{\Q}$  and $ \U(1)_{\L}$ 
present at the renormalizable level,\footnote{\label{foot:CP}
The new sector also respects a $\mathcal{C}$ symmetry~\cite{Buttazzo:2019mvl} defined as $\S\to \S^*$, $\mathcal{D_\mu}\to\mathcal{D}_\mu^*$, $\mathcal{F} \to i\gamma^2 \mathcal{F}^\dagger$, with $\mathcal{F}$ denoting the various fermions. On the $\Sp(\N)$ bosonic multiplets, $\mathcal{C}$ acts as follows
\beq
s \to s ,\qquad
 \ a \to -a ,\qquad \ \tilde{s}\to -\gamma_\N\tilde{s}\gamma_\N ,\qquad
 \mathcal{A}\to -\gamma_\N\mathcal{A}\gamma_\N ,\qquad
  \mathcal{W}\to \gamma_\N\mathcal{W}\gamma_\N \,.
\eeq
The colored fermions $\mathcal{\Q}$ relate this symmetry to the SM via QCD, so that $\mathcal{C}$ (combined with parity) extends the usual charge-parity conjugation CP, under which the axion is odd, as it should.} 
we will later discuss 
how the situation changes at the non-renormalizable level.

\subsection{Axion effective Lagrangian}
\label{sec:EFTaxion}
Under a $\U(1)_{\text{PQ}}$ rotation with angle $\alpha$ the fields transform as 
\begin{align*}
a\rightarrow a+\alpha \sqrt{\N/2}w ,\qquad
\Q \rightarrow e^{+i\alpha/2}\Q,\qquad
\L^i \rightarrow e^{-i\alpha/2}\L^i .
\end{align*}
The QCD and QED anomalies are 
\beq
\mathcal{A}_{\rm QCD}=\N\frac{g_3^2}{32\pi^2} G^{a\mu\nu}\Tilde{G}^a_{\mu\nu},\qquad
\mathcal{A}_{\rm QED}=3\N\frac{e^2}{16\pi^2}(Y_\Q^2-Y_{\L}^2) F^{\mu\nu}\Tilde{F}_{\mu\nu} .
\eeq
In view of the above anomalies the effective axion Lagrangian is 
\begin{align}
\mathcal{L}_{\text{anom}}^{\text{eff}}&=\frac{a}{\sqrt{\N/2}w}\N\frac{g_3^2}{32\pi^2}G^{a\mu\nu}\Tilde{G}^a_{\mu\nu}+
\frac{a}{\sqrt{\N/2}w}3\N(Y_\Q^2-Y_{\L}^2)\frac{e^2}{16\pi^2}F^{\mu\nu}\Tilde{F}_{\mu\nu} \nonumber \\
&\equiv\frac{a}{f_a}\frac{g_3^2}{32\pi^2}G^{a\mu\nu}\Tilde{G}^a_{\mu\nu}+
\frac{g_{a\gamma}^0}{4}a F^{\mu\nu}\Tilde{F}_{\mu\nu},
\label{eq:effaxionanomaly}
\end{align}
where in the last step we identified the axion decay constant $f_a=w/\sqrt{2\N}$
and coupling to photons $g_{a\gamma}^0=\sfrac{3e^2(Y_Q^2-Y_{\L}^2)}{4\pi^2 f_a}$. 
After rotating away the $aG\tilde G$ term via an axion-dependent light quark field 
redefinition, the axion-photon coupling gets dressed via the axion-pion mixing as  
$g_{a\gamma}= \alpha_{\text{em}} C_{a\gamma} / (2\pi f_a)$
in terms of the dimension-less coupling \cite{Kaplan:1985,Srednicki:1985,Villadoro:2016}
\beq 
\label{eq:Cagamma}
C_{a\gamma} = 6(Y_\Q^2-Y_\L^2)-1.92(4) .
\eeq

\subsubsection*{Axion Domain Walls}
The QCD anomaly breaks $\U(1)_{\rm PQ} \to \mathbb{Z}_{\N}$, 
since the $2\pi f_a$-periodic
axion potential has $\N$ degenerate minima when the axion field is 
varied in its angular domain 
$a \in \[0, 2\pi\) \sqrt{\N / 2} w = \[0, 2\pi\) \N f_a $. 
However, the $\mathbb{Z}_{\N}$ action can be embedded in the $\SU(\N)$ center, 
thus making the axion minima gauge equivalent. This avoids the formation 
of axion domain walls at the QCD phase transition, 
and solves the axion domain wall 
problem along the lines of \cite{Lazarides:1982tw,DiLuzio:2017tjx}.


\subsection{Higher dimensional operators}\label{PQpotential}
So far we considered the renormalizable theory.
As anticipated, the relevance of the present model consists 
in the fact that 
the PQ symmetry arises
accidentally at the renormalizable level and can remain good enough even in the presence of possible non-renormalizable operators. 
In this section we study how effective operators can break accidental symmetries, 
hence contributing to the axion potential and to the decay of heavy relics. 
 

\subsubsection*{PQ-breaking operators}
We defined the PQ symmetry such that the PQ charge of any field is proportional to its $\N$-ality
(number of lower indices minus number of higher indices).
This means that operators containing one 
$\epsilon_\N$ 
tensor of $\SU(\N)$ break the PQ symmetry,
while it is preserved by all other operators,
such as the renormalizable Yukawa couplings of eq.\eq{YukSp}, or the dimension-7 operator $(q_R \Q_L) \S^* (q'_R \Q_L )$. 
Note, also, that thanks to the $\N$-ality selection rule, 
dimension-7 operators are guaranteed to preserve the PQ symmetry even 
in the presence of additional ``quarks'' and ``leptons'' in arbitrary $\SU(3)_c \times \U(1)_Y$ 
representations.

We then search for the lowest-dimensional operator containing one $\epsilon_\N$ tensor.
This is built by contracting with scalars $\S$, as 
replacing one $\S_{IJ}$ with two fermions $\Q_I$ or $\bar \L_I$ increases the dimension of the operator.
The lowest dimensional operator that explicitly breaks the accidental PQ symmetry is
the Pfaffian, with dimension $\N/2$
\beq \label{eq:Pf}
 {\rm Pf}\,\S =\sqrt{\det\S}=\epsilon_{\N}^{I_1I_2\dots I_{\N-1}I_{\N}}\S_{I_1I_2}\dots \S_{I_{\N-1}I_{\N}}.\eeq
Assuming that new physics generates such operator with coefficient suppressed by some scale $\Lambda_{\rm UV}$, its contribution to the axion potential originates from
\begin{equation}\label{detpotential}
\frac{e^{i \varphi}}{\Lambda_{\rm UV}^{\N/2-4}} {\rm Pf}\,\S + \hbox{h.c.} , 
\end{equation}
where $\varphi$ is a generic CP-violating phase. 
Inserting ${\rm Pf}\,\S = w^{\N/2} e^{i a/2f_a} + \cdots $ 
the axion potential obtained from QCD plus eq.~\eqref{detpotential} is
\beq
V_a=-m_\pi^2f_\pi^2\cos\left( \frac{a}{f_a}+\bar{\theta}\right)+\frac{2 w^{\N/2}}{\Lambda_{\rm UV}^{\N/2-4}}\cos\left(\frac{a}{2f_a}+\varphi\right), 
\eeq
where $\bar{\theta}$ is the QCD topological term 
in a basis in which the SM quark masses are real.
The experimental bound $\langle a/f_a \rangle+\bar{\theta}<10^{-10}$ 
is satisfied for
\beq
f_a\lesssim \frac{\Lambda_{\rm UV}}{\sqrt{\N/2}}\left(\frac{m_\pi f_\pi}{\Lambda_{\rm UV}^2}\right)^{4/\N}\times 10^{-20/\N}.
\label{eq:facondition}
\eeq
Note that this bound holds even if the new physics respects CP, $\varphi=0$,
as the operator would not relax the axion to the field value that cancels CP violation at low energy
and hence it would not cancel the $\bar\theta$ term.
If $\Lambda_{\rm UV} \sim M_{\rm Pl}$, the phenomenological bound $ f_a \gtrsim 10^{11} \,\text{GeV} $ for PQ symmetry broken after inflation requires $\N\gtrsim 24$. 
If, instead, the PQ symmetry is broken before inflation
$f_a$ can be as low as $10^9$ GeV and then 
 $\N\gtrsim 20$ suffices.
In our numerical plots we will assume for definiteness a Planckian cut-off.
If instead $\Lambda_{\rm UV} \approx 2\times10^{16}\GeV$ (as motivated e.g.~by supersymmetric unification) one needs
$\N\gtrsim24$ for $f_a \gtrsim 10^{9} \,\text{GeV}$.

The quality of the PQ symmetry is strengthened by the following argument.
The operator in eq.~(\ref{detpotential}) can have a coefficient 
$\Lambda_{\rm UV}\ll M_{\rm Pl}$ if it is mediated by 
particles with renormalizable couplings below the Planck scale.
One then needs a Yukawa or a scalar quartic containing the $\epsilon_\N$ tensor,
which implies particles in a $\SU(\N)$ representation with
$n\sim \N/2$ or $n \sim \N/3$ indices. The resulting large
contribution to $\beta(\g) \sim \N^n$ implies that such particles cannot
be much below the Planck scale.

\subsubsection*{$\Q$-decay operators}
Furthermore, gauge invariance allows dimension 6 operators such as
\beq\label{Qdecayoperators}
(q_R \Q^{I}_{L})(e_R \L_I) ,\qquad (q_R \Q^I_{R})(q'_R \L_I),
\eeq
(where $q_R = (u_R, d_R)$ and $e_R$ are left-handed Weyl spinors, $\SU(2)_L$-singlet  
SM quarks and leptons)
that conserve the PQ symmetry and
break the other accidental global U(1)'s as
\beq  \U(1)_\Q\otimes\U(1)_\L  \to \U(1)_{\Q-\L}\eeq
such that only the lightest state containing $\Q$ and/or $\L$ is stable.
Assuming that $\Q$ is heavier than $\L$, 
heavier Sp bound states containing $\Q$ decay with rate \cite{Asatrian:2012tp}
\begin{equation}\label{eq: Qrate}
\Gamma_\Q\approx\frac{m_\Q^5}{4(4\pi)^3 \Lambda_{\rm UV}^4}\left[\frac{1-x^2}{2}+x\ln x\right]
\approx \frac{1}{13\,{\rm sec}} \left(\frac{m_\Q}{2\times 10^{11} \GeV}\right)^5\left(\frac{M_{\rm Pl}}{\Lambda_{\rm UV}}\right)^4, 
\end{equation}
for $x\equiv m_\L/m_\Q\simeq 0.5$.
We can reasonably approximate the Big Bang Nucleosynthesis (BBN) bound on $\Q$ decays~\cite{astro-ph/0408426} by simply demanding that $\tau_\Q = 1/\Gamma_\Q < 0.1\,{\rm sec}$ such that $\Q$ decays before BBN.
This gives the bound
\beq\label{eq:fabound}
f_a\gtrsim  \frac{1}{y_\Q} \sqrt{\frac{10}{\N}}\left(\frac{\Lambda_{\rm UV}}{M_{\rm Pl}}\right)^{\sfrac{4}{5}}
\times\left\{\begin{array}{ll}
1.2\times  10^{11}\GeV & \hbox{for $x=1/2$}\\
0.7\times  10^{11}\GeV & \hbox{for $x \ll 1$}
\end{array}\right.  .
\eeq
Furthermore we neglect the possibility that $\Q$ decays while dominating the energy density of the universe
and thereby providing extra reheating and dilution.

\subsubsection*{$\L\L$-decay operators}
The Sp mesons $\L\gamma_{\N}\L$ with mass
\beq 
\label{eq:MLL}M_{\L\L} \sim \max (\Lambda_{\Sp},2y_\L w),
\eeq
are possibly stable but are charged if $Y_\L\neq 0$.
We avoid heavy charged relics assuming 
\beq 
\label{eq:YQdecay}
Y_\L=0 \quad \text{and} \quad Y_\Q=\{-\frac{1}{3}, \frac{2}{3}, -\frac{4}{3}\} .
\eeq
For $Y_\L=0$  the Sp mesons are kept stable only by the accidental flavour symmetry
that arises at renormalizable level and that gets broken if dimension-6 operators such as
$\L^i \S\S^\dagger \S\L^j$ have a different flavour structure.
One then expects that $\L\gamma_{\N}\L$ mesons decay fast,
leaving no relics despite that $\L$ is stable.
We will however also mention the alternative possibility that $\L\gamma_\N\L$ is an acceptable DM candidate with
$\tau_{\L\L}\circa{>} 10^{26}\,{\rm sec}$, given that flavour
couplings might be small and not approximated by a single scale $\Lambda_{\rm UV}$.



\begin{figure}[t]
\centering
$$\includegraphics[width=0.45\textwidth]{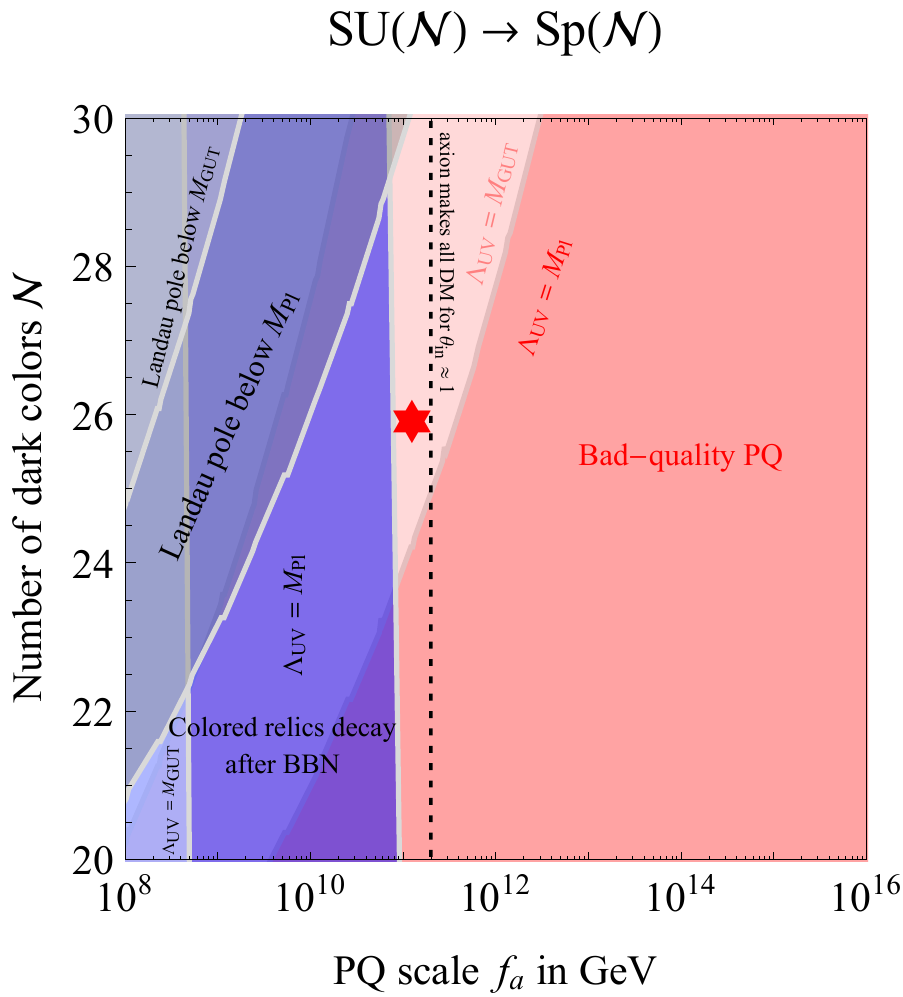} \qquad
\includegraphics[width=0.45\textwidth]{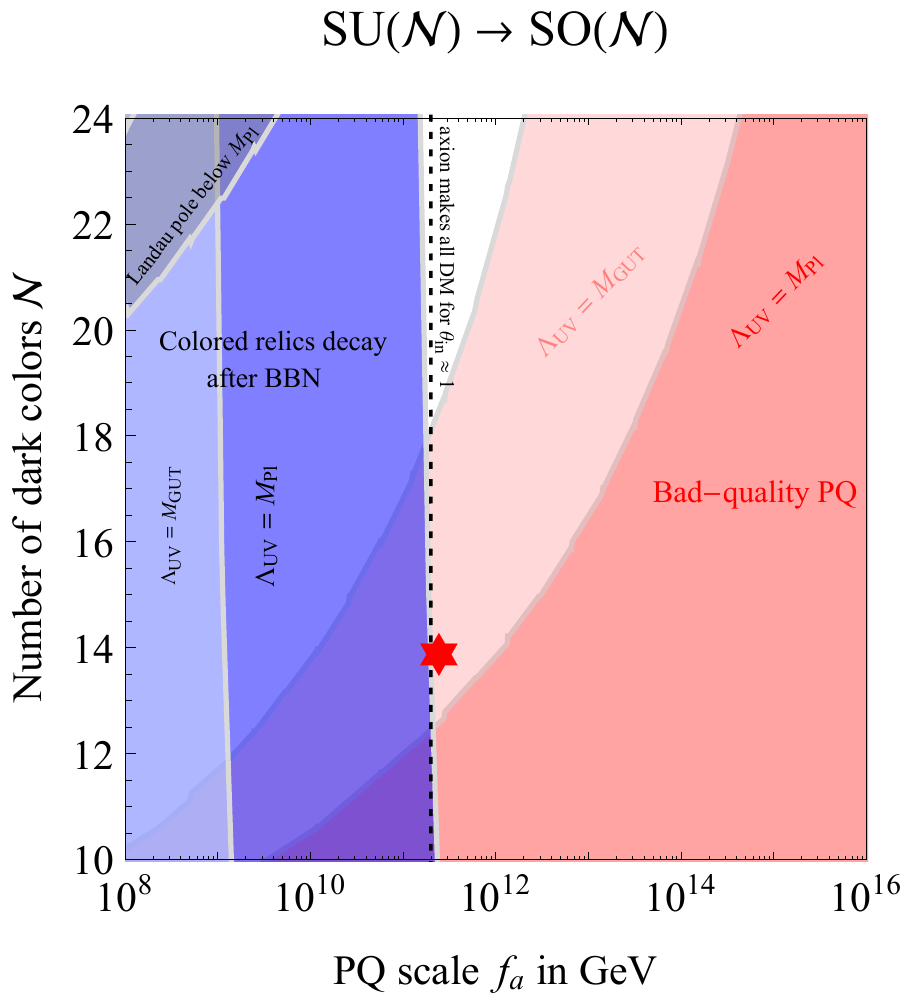}$$
\vspace{-1cm}
\caption{\em \label{fig:tauQasymm}
Values of $(f_a, \N)$ such that:
i) the model gives rise to a high PQ-quality axion (the region shaded in red is excluded);
ii) does not have sub-Planckian Landau poles for $g_3$  (gray is excluded);
iii) colored relics decay before BBN (blue is excluded if PQ is broken after inflation).
We assumed $y_\Q=1 = 2 y_\L$ and non-renormalizable operators suppressed by $\Lambda_{\rm UV}\approx M_{\rm Pl}$
(darker regions) or $M_{\rm GUT}\approx 2~10^{16}\GeV$ (lighter regions).
$\N$ should be understood as an integer parameter 
(even integer for the left-hand plot).
{\bf Left}: model with a scalar in the anti-symmetric. 
{\bf Right}:  model with a scalar in the anti-symmetric.
The stars indicate the models considered in fig.\fig{RelicPreInf},\fig{RelicPostInfSO},
that satisfy all the bounds.
They lie around the vertical dashed line, where
axions with initial $\theta_i \sim 1$ make all DM.
}

\end{figure}

%
The bounds related to Landau poles 
and higher-dimensional operators (from BBN and PQ quality)
are plotted in fig.\fig{tauQasymm}, which shows
that they can all be simultaneously satisfied for $\N\circa{>}24$ 
and physically acceptable values of the axion decay constant $f_a$.

\subsubsection*{Axion-photon coupling predictions}

\begin{figure}[t]
\centering
$$\includegraphics[width=1.0\textwidth]{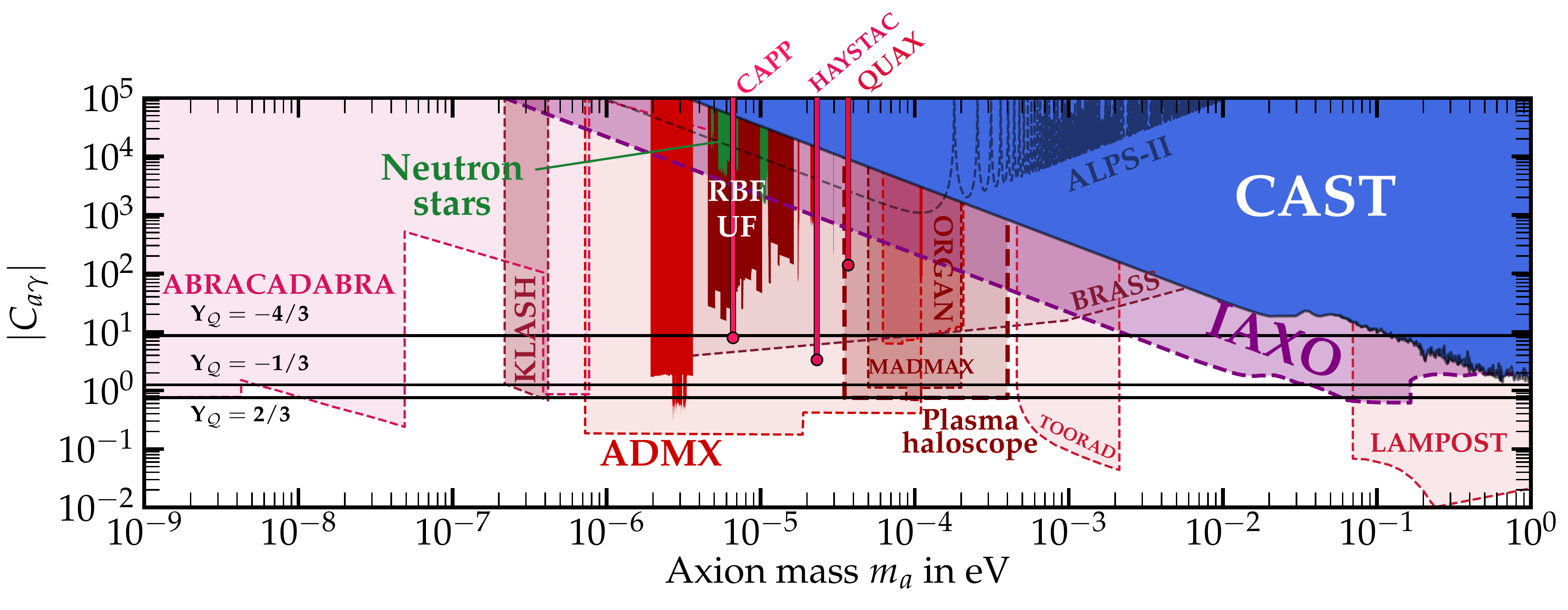}$$
\vspace{-1cm}
\caption{\em \label{fig:axionphoton}
Predictions for the axion-photon coupling in eq.\eq{Cagamma}
and sensitivity of present 
and future axion experiments. Axion limits from \cite{ohare}.}
\end{figure}

Requiring that the colored exotic states $\Q$ decay fast enough 
to avoid problems with cosmology
allows to fix their hypercharges and in turn to predict the axion-photon 
coupling, following a similar strategy as in the case of  
KSVZ axions \cite{DiLuzio:2016sbl,DiLuzio:2017pfr}. 
In particular, in fig.\fig{axionphoton}  
we show the predictions for the dimension-less axion-photon 
coupling in eq.~(\ref{eq:Cagamma}) according to the 
values of the hypercharges in eq.~(\ref{eq:YQdecay}). 
Current limits (full lines) and projected ones (dashed lines) 
from axion experiments 
are displayed as well. 
It should be noted that most of those experiments  
(apart for CAST, IAXO and ALPS-II)
assume that the axion comprises the 100$\%$ of DM. 
Consequently their sensitivity is diluted as $(\Omega_a / \Omega_{\rm DM})^{1/2}$, 
if the axion is only a fraction of the whole DM. 
Hence, we next study the cosmology and the composition of the DM abundance.  

\subsection{Cosmology and Dark Matter}\label{cosmo}
The model contains two DM candidates: the axion condensate and possibly the $\L\gamma_{\N}\L$ meson if $Y_\L=0$.
The reheating temperature after inflation is $T_{\rm RH} \sim\sqrt{M_{\rm Pl}H_{\rm infl}}$ if reheating
happens instantaneously, or $T_{\rm RH} \sim\sqrt{M_{\rm Pl}\Gamma_{\rm infl}}$ otherwise.
Qualitatively different cosmological histories arise depending on whether $T_{\rm RH}$ is high 
enough so that the 
PQ symmetry is restored after inflation.

\subsubsection{PQ broken before inflation, $T_{\rm RH}\circa{<} f_a$}\label{Spbefore}
If the PQ symmetry is broken during or before inflation and is not restored afterwards,
the Hubble rate during inflation must be smaller than
\beq H_{\rm infl} \lesssim 10^8\,\text{GeV} \frac{\theta_i}{\pi} \frac{\Omega_{\rm DM}}{\Omega_a} \frac{f_a}{10^{12}\,\text{GeV}}\eeq
in order to avoid excessive axion  iso-curvature fluctuations during inflation
(see e.g.~section 3.5 of~\cite{2003.01100}).
This implies an upper bound on $T_{\rm RH}$, that  anyhow must be smaller than $f_a$ under the present assumptions.

\begin{figure}[t]\centering
$$\includegraphics[width=0.40\textwidth]{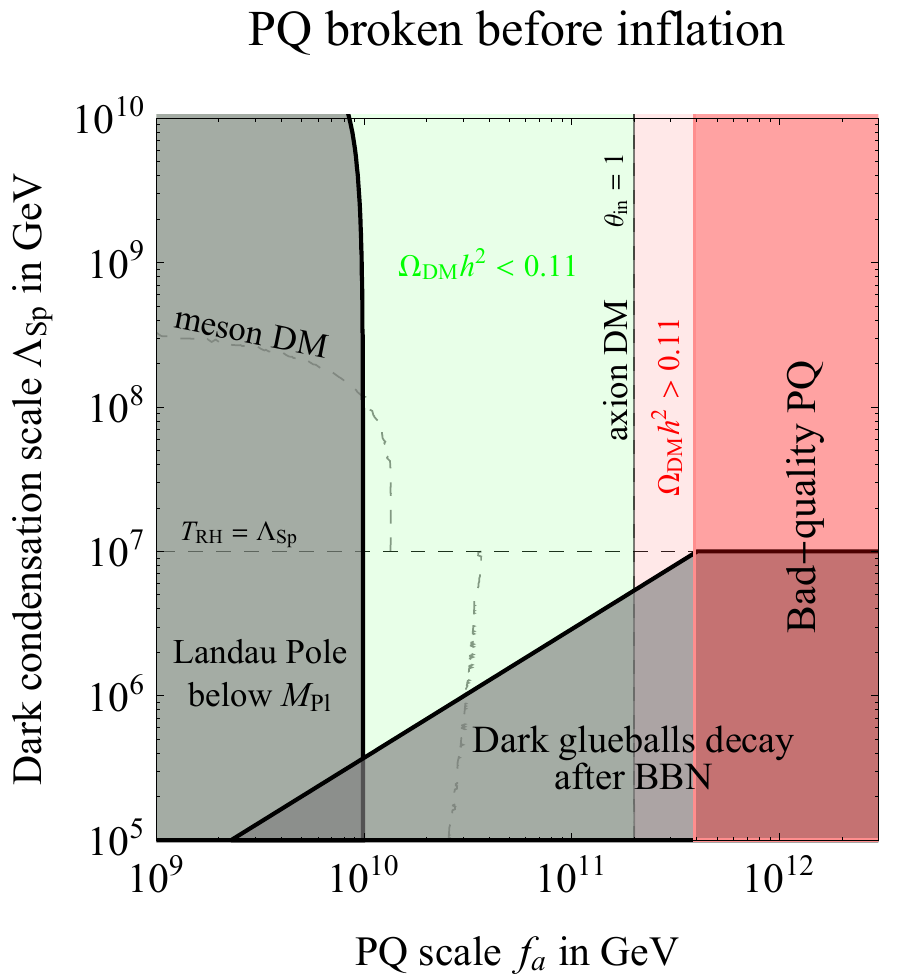}\qquad
\includegraphics[width=0.40\textwidth]{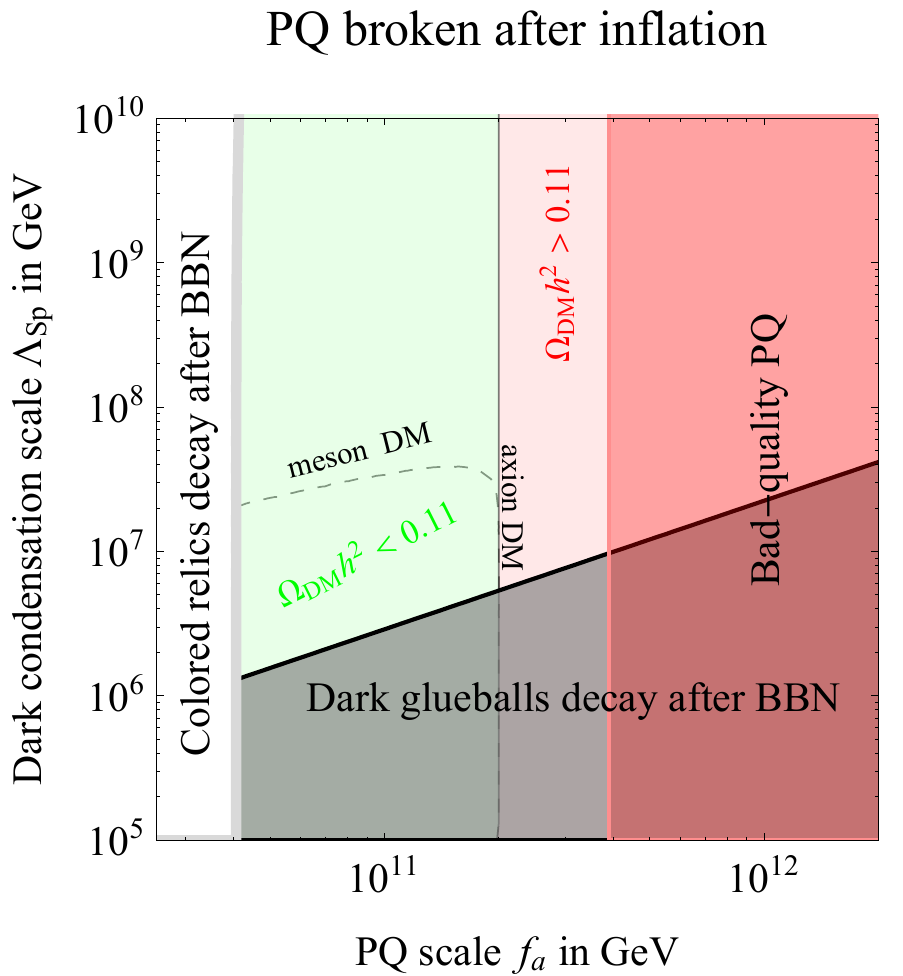}$$
\vspace{-1cm}
\caption{\em\label{fig:RelicPreInf}\label{fig:RelicPostInf} 
Parameter space of the $\SU(26)$ model assuming
$Y_\L=0$, $Y_\Q=-1/3$, 
and $y_\Q=1$. 
We chose a large $\N$ such that the cosmological DM abundance is reproduced along the red/green boundary
(the dashed curve shows how this boundary would change if, in addition to the axion,
a cosmologically stable $\L\L$ meson with $y_\L=10^{-3}$ contributes to DM;
otherwise the same plots apply for any $y_\L \ll y_\Q$).
The right red region is disfavoured by the PQ quality arguments in eq.\eq{facondition}. 
{\bf Left:} assuming that the PQ symmetry is broken before inflation with $T_{\rm RH} = 10^7\GeV \ll f_a$.
The left gray region is excluded by sub-Planckian Landau poles.
{\bf Right:} assuming that $T_{\rm RH}\circa{>}f_a$, i.e.\ PQ-breaking after inflation.
The left white region is excluded because colored relics decay after BBN ($\tau_\Q>0.1\,{\rm sec}$). 
}
\end{figure}

All the heavy stable relics including topological defects (strings from PQ breaking)
get diluted during the inflationary expansion. 
The abundance of the DM candidates is estimated as follows:
\begin{itemize}
\item The axion DM abundance produced through the misalignment mechanism 
can be analytically approximated as \cite{2003.01100}
\begin{equation}\label{eq:axionabundance}
\frac{\Omega_a h^2}{0.12} \approx \theta_{\rm in}^2\left(\frac{f_a}{2.0\times10^{11}\,\text{GeV}}\right)^{7/6} , 
\end{equation}
where the initial axion phase $\theta_{\rm in}$ is expected to be of order one, but can accidentally be smaller.

\item The $\L\gamma_{\N}\L$ meson $\M_\L$ with mass $M_{\L\L}$ given in eq.\eq{MLL} might be stable and
light enough that it is produced thermally.
\begin{itemize}

\item If $T_{\rm RH}\circa{>} \Lambda_{\rm Sp}$ and possibly larger than $m_\L$,
its constituents can be produced
from $\A\A\to\L\bar \L$  with rate $\gamma \sim \g^4 T^4 e^{-2m_{\L} /T}$.
The resulting number abundance is
\beq \label{eq:YLL1}
Y_{\L\L} \approx  \max_T \frac{\gamma}{H s}  \sim   \frac{ \g^4M_{\rm Pl} }{\min (m_\L,T_{\rm RH})}
e^{- 2 m_{\L}/T_{\rm RH}}\eeq
and gets later diluted by glue-ball decays.

\item If $T_{\rm RH}\circa{<} \Lambda_{\rm Sp}$ 
it can be produced
from $a X \to \M_\L \bar \M_\L$
with space-time density rate 
$\gamma \sim \Lambda_{\rm Sp}^4  T^4 e^{-2M_{\L\L}  /T}/f_a^4$.
The resulting number abundance is
\beq \label{eq:YLL2}
Y_{\L\L} \approx \left.\frac{\gamma}{H s}\right|_{T =T_{\rm RH}}  \sim e^{-2 M_{\L\L}/T_{\rm RH}} \frac{M_{\rm Pl} \Lambda_{\rm Sp}^4}{T_{\rm RH}f_a^4}.\eeq
\end{itemize}
\end{itemize}
In fig.\fig{RelicPreInf}a we show the parameter space of the model for some 
representative benchmark values. 
We select a  high $\N=26$ such that DM can be composed solely by axions:
in view of eq.\eq{axionabundance} this happens along the vertical line in fig.\fig{RelicPreInf}.
More likely $\L\L$ are unstable and the value of $y_\L$ is irrelevant.
We assumed a small $y_\L=10^{-3}$ such that,
if $\L\L$ is stable, a second DM branch appears in fig.\fig{RelicPreInf}a
in which DM is composed by mesons via thermal production.
The transition between eq.\eq{YLL1} and eq.\eq{YLL2} gives rise to the discontinuity at $T_{\rm RH}\sim \Lambda_{\rm Sp}$
in fig.\fig{RelicPreInf}a.

%

\subsubsection{PQ broken after inflation, $T_{\rm RH}\circa{>} f_a$}\label{Spafter}
We here consider the alternative situation where 
the reheating temperature is high enough that the PQ symmetry is restored after inflation
and all the new states predicted by the model thermalise. 

Demanding that axions do not exceed the cosmological DM density gives the bound
$f_a < 2.0\times10^{11}\,\text{GeV}$ for 
the average initial misalignment angle 
$\theta_{\rm in} = 2.2$. 
The latter numerical result
(more precise than eq.~\eqref{eq:axionabundance})
is obtained by tracking the temperature dependence  of the topological susceptibility via lattice QCD simulations \cite{Borsanyi:2016ksw}.
If the PQ symmetry is broken after inflation, topological defects 
(strings and domain walls)
add up to the total axion relic density. 
The contribution of would-be disastrous axion domain walls to the energy density 
is avoided thanks to the mechanism outlined at the end of section \ref{sec:EFTaxion},  
while that of axion strings is relevant, but difficult to be estimated 
(see e.g.~\cite{Gorghetto:2018myk}). 
We neglect it here for simplicity, also because a complete consensus on 
their 
importance has not been achieved yet in the literature. 
However, recent developments indicate that they would strengthen 
the upper bound on $f_a$ by more than one order of magnitude \cite{Gorghetto:2020qws}.

On the other hand, the possibly stable DM candidate $\L\gamma_{\N}\L$ risks being over-abundant.
Its thermal relic abundance, $Y_{\L\L}\sim1/T_\text{dec}M_\text{Pl}\sigma_\text{ann}$ with $\sigma_\text{ann}\sim1/\Lambda_{\Sp}^2$ and $T_\text{dec}\sim\Lambda_{\Sp}$
(section 2.3 of~\cite{1907.11228} contains an extended discussion),
is over-abundant if its mass $\MLL$
is above than $100\TeV$, the critical value above which even strongly-coupled freeze-out leaves too much DM.
A  $\L\gamma_{\N}\L$ lighter than 100 TeV needs a very small $y_\L$ and $\Lambda_{\rm Sp}$.

Furthermore, the colored relics and the glue-balls of $\Sp(\N)$
must decay before BBN, but their decay can be 
slow enough so that they substantially reheat the universe.
Dark glue-balls decays dilute all relics containing $\Q$ and $\L$ heavy fermions (while the axion density is still 
a cosmological constant and hence is not diluted),
allowing us to get a multi-component 
$\L\gamma_{\N}\L$ plus axion abundance that matches the DM relic density. 
We anyhow demand that colored relics decay before BBN, which implies the bound found in eq.~(\ref{eq:fabound}).

\begin{figure}[t]
\centering
$$\includegraphics[width=0.3\textwidth]{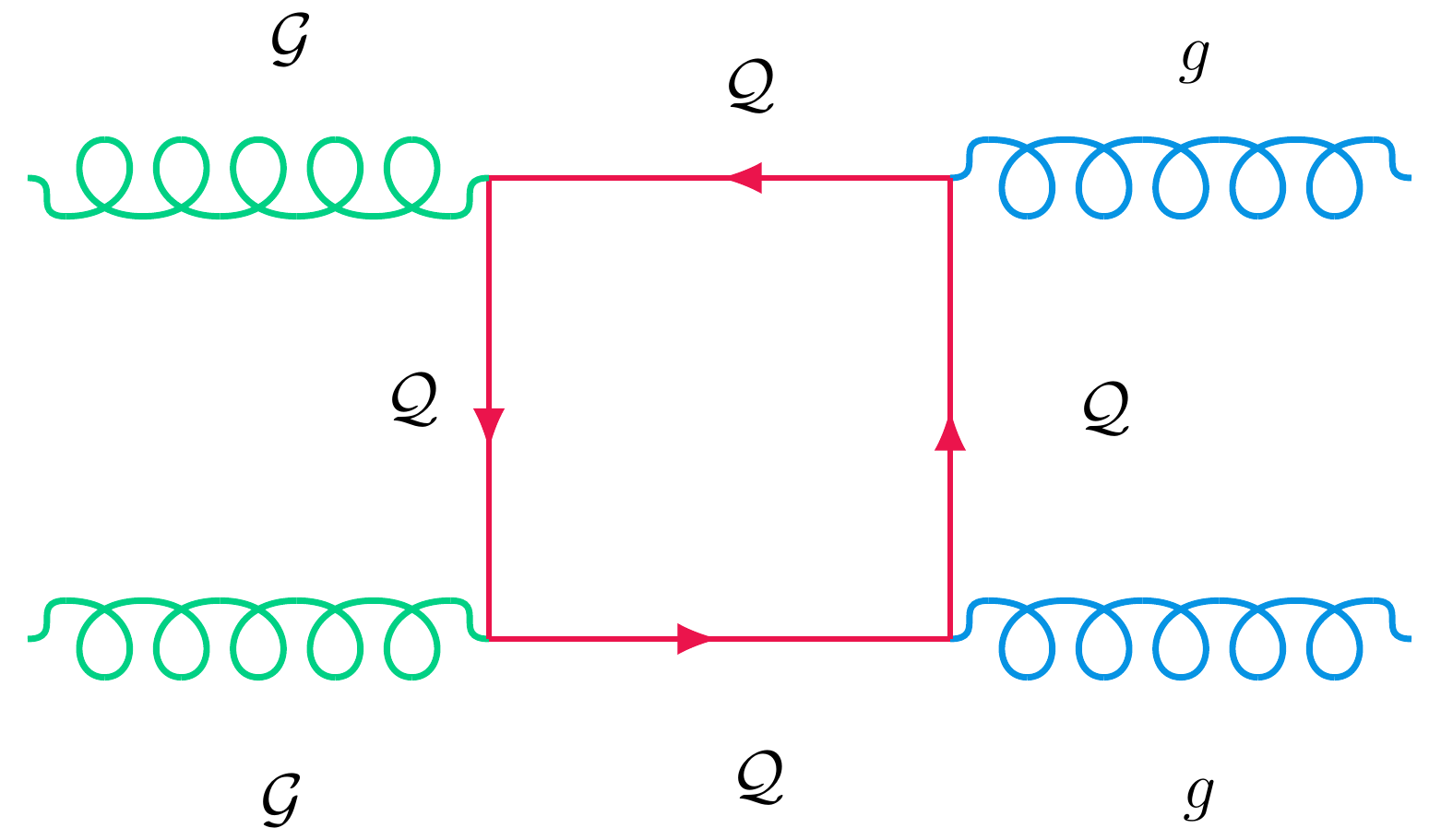}\qquad 
\includegraphics[width=0.3\textwidth]{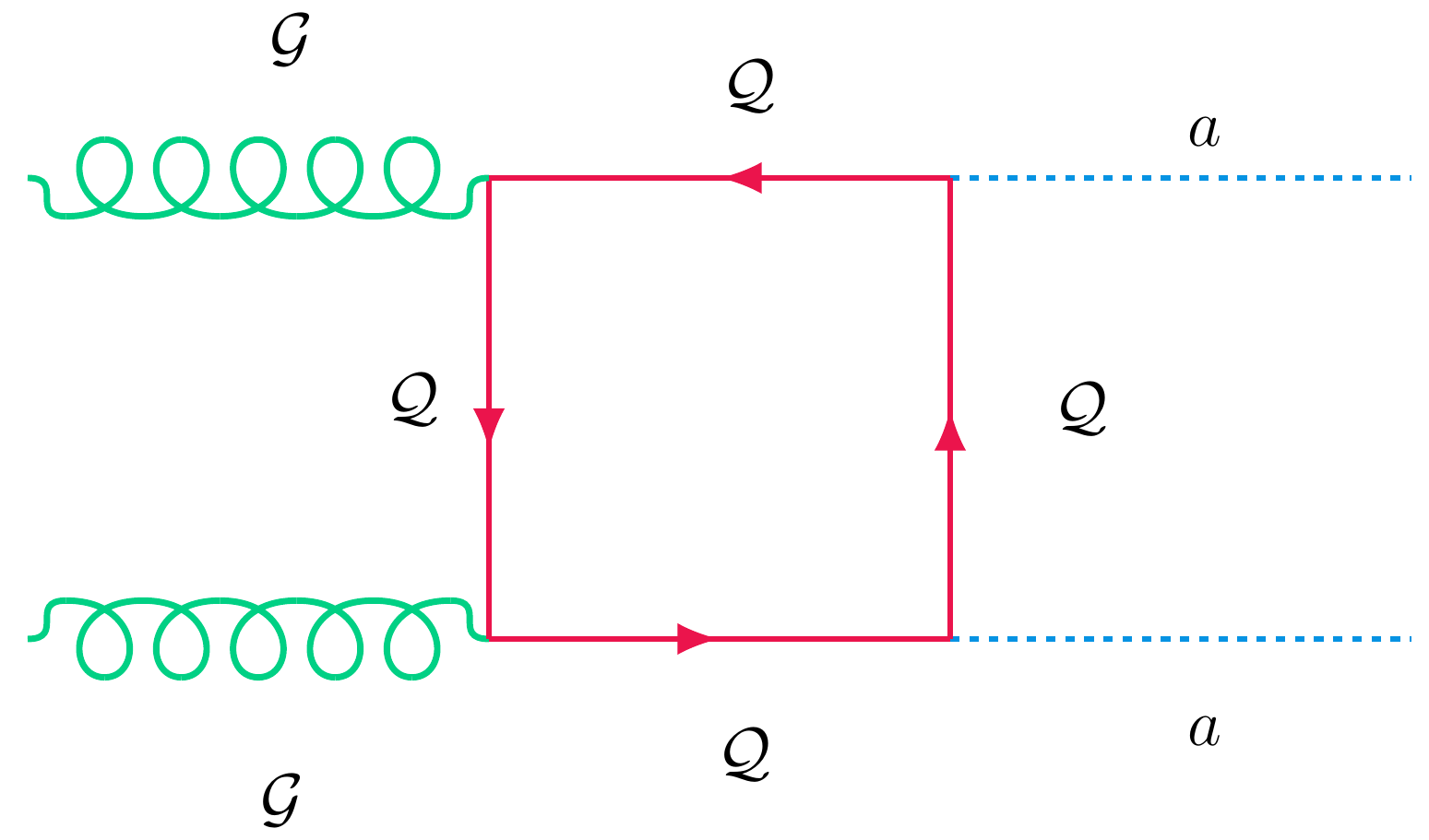}\qquad 
\includegraphics[width=0.3\textwidth]{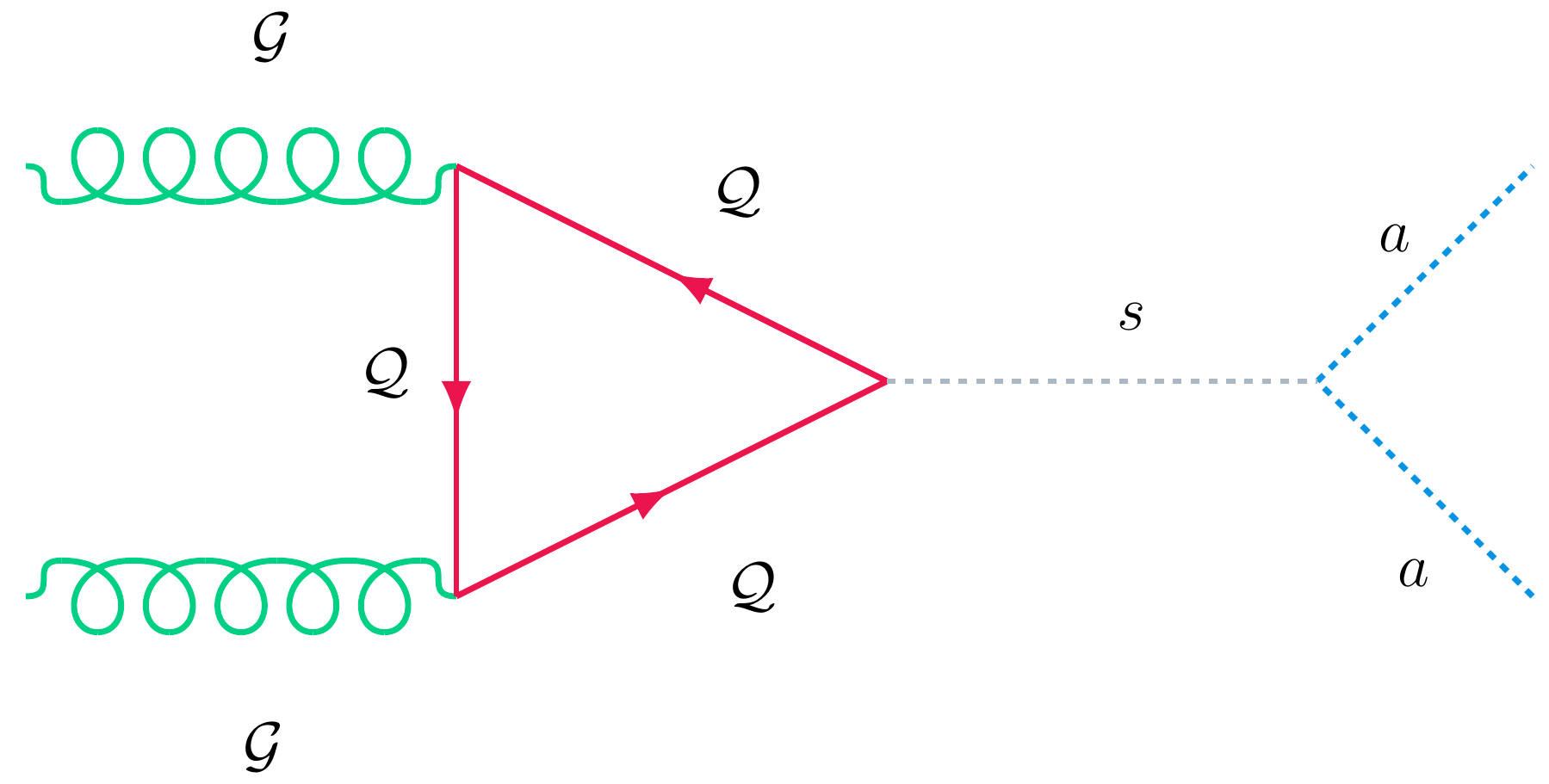}$$
	\vspace{-1cm}
	\caption{\em \label{fig:Feyn}
Some Feynman diagrams for dark glue-ball decay into gluons and axions.
Analogous diagrams with heavy leptons $\mathcal{L}$ or vectors $\mathcal{W}$ in the loops are not plotted.}
\end{figure}

Glue-balls with mass $\sim \Lambda_{\rm Sp}$ that decay at $T =T_{\rm decay} \sim(\Gamma_{\rm DG}^2 M_{\rm Pl}^2/\Lambda_{\rm Sp})^{1/3}$ reheat the universe up to
$T_{{\rm RH}'} \sim  T_{\rm decay} (\Lambda_{\rm Sp}/T_{\rm decay})^{1/4}$.
The dilution factor is estimated as\cite{Landini:2020}
\begin{equation}\label{dilu}
D^{-1}=\left[ 1+\frac{g_{{\rm DG}}}{g_\text{SM}^{2/3}}\left(\frac{\Lambda_{\Sp}^2}{\GammaDG M_\text{Pl}}\right)^\frac{2}{3} \right]^{3/4},
\end{equation}
where $g_{\rm SM} \approx 100$ is the number of SM degrees of freedom, and
$g_\text{DG}=\N(\N+1)$ is the number of Sp dark gluons~\cite{1707.05380}.
A large $D$ can need a baryogenesis mechanism below $T_{{\rm RH}'}$;
we do not address such model-dependent issues.

Dark glue-balls with mass  $M_\text{DG}\approx7\Lambda_{\Sp}$  can decay into axions and SM particles
through the Feynman diagrams in fig.\fig{Feyn}.
The rates are estimated as follows:
\begin{itemize}
\item Glue-balls decay into gluons $g$ through a loop of $\Q$ (left-handed diagram in fig.\fig{Feyn})
that gives the dimension 8 effective
operator $\A_{\mu\nu}^2 g_{\alpha\beta}^2$ (see eq. (41) of~\cite{1707.05380})
\beq  \Gamma_{\text{DG}\to gg}  \sim \frac{\alpha_{\rm DC}^2\alpha_3^2 M_\text{DG}^9}{m_\Q^8}.\eeq

\item Glue-balls decay into axions through a loop of $\Q$ or $\L$ (middle diagram in fig.\fig{Feyn})
that gives the dimension 8 operator $\A_{\mu\nu}^2 (\partial_\alpha a)^2$
\beq \label{eq: DG-decaytoaa1}  \Gamma_{\text{DG}\to aa}  \sim \frac{ \alpha_{\rm DC}^2 M_\text{DG}^9}{m_{\Q,\L}^8}\left(
 \frac{m_{\Q,\L}}{4\pi f_a} \right)^2.\eeq
A similar diagram gives decays into $ag$.

\item Glue-balls decay into axions as described by the right-handed diagram in fig.\fig{Feyn}: 
through the dimension 5 operator $(18-7\N)\alpha_{\rm DC}\A_{\mu\nu}^2 s/16\pi w $ times the dimension 5 
$s (\partial_\alpha a)^2$ operator 
\beq\label{eq: DG-decaytoaa2} \Gamma_{\text{DG}\to aa}\approx \frac{(18-7\N)^2\alpha^2_{\rm DC} M_\text{DG}^9}{512\pi^3M_s^4 f_a^4}.\eeq

\item Glue-balls decay into $H H^\dagger =\{hh, hZ,ZZ,W^+W^-\}$ proceed through the coupling $\lambda_{HS}$
that connects the dark and the SM sectors (that however can be small, in order to avoid an unnaturally large
contribution to the Higgs mass). Taking into account that $h$ mixes with $s$ we obtain, for $M_{\rm DG}\gg M_{h,W,Z}$
\begin{equation}
\Gamma_{\text{DG}\to H H^\dagger} \approx \frac{(18-7\N)^2\alpha^2_{\rm DC}\lambda_{H\S}^2  M_\text{DG}^5}{2048\pi^3 M_s^4}.
\end{equation}
\end{itemize}
The most important channels are \eqref{eq: DG-decaytoaa1} and \eqref{eq: DG-decaytoaa2} 
for $y_\L\ll1$, while \eqref{eq: DG-decaytoaa2} dominates if $y_\L\sim1$.

The parameter space is plotted in fig.\fig{RelicPostInf}b: we see that regions exist where
all constraints are satisfied, the axion quality is good, and the DM cosmological abundance is reproduced
either through axions or through dark mesons (if stable), or both.

\section{Symmetric scalar that breaks $\SU(\N)\to\SO(\N)$}\label{SO}
We next consider a different but similar model: the scalar is now in the symmetric representation
and spontaneously breaks the $\SU(\N)$ gauge group to $\SO(\N)$.
We avoid repeating the many aspects of the model which remain similar to those in section~\ref{Sp}
and highlight the key differences. 

If $\N>4$ the most generic renormalizable Lagrangian has the same form as in eq.s~\eqref{lagrangian} and \eqref{sys:Lags}.
The Yukawa matrix $y_\L$ is now symmetric and can be rotated to a basis where it is diagonal with real positive entries. 
Unlike the anti-symmetric scalar, where the gauge invariant 
Pfaffian operator is defined only for even $\N$, 
the symmetric scalar model can also be constructed using odd values of $\N$.
We again require that couplings do not hit Landau poles below the Planck scale;
SM couplings run as in the previous model, giving the same constraints.



\subsubsection*{Accidental symmetries}
The situation is similar to the model of section~\ref{Sp}, but with two differences.

First, in both models the $\SU(\N)$ theory is accidentally invariant under a reflection,
which we dub U-parity \cite{Buttazzo:2019mvl}, 
of any of the $\N$ equivalent directions in group space.
U-parity is obtained by flipping the sign of any color, for example the 1st one.
This flips the signs of those generators with an $1I$ entry,
preserving the $\SU(\N)$ Lie algebra, such that U parity acts on
components of vectors in the adjoint and of other multiplets
as
\be\label{eq:Uparity}
\G_I^J \stackrel{\P_\text{U}}\to (-1)^{\delta_{1I} + \delta_{1J}} \G_I^J,
\qquad 
{\Q}_{I} \stackrel{\P_{\rm U}}\to (-1)^{\delta_{1I}}  {\Q}_{I} ,\qquad
{\L}_{I} \stackrel{\P_{\rm U}}\to (-1)^{\delta_{1I}}  {\L}_{I} ,\qquad
{\cal S}_{IJ} \stackrel{\P_{\rm U}}\to (-1)^{\delta_{1I}+\delta_{1J}}  {\cal S}_{IJ} 
\ee 
having written the $\SU(\N)$ vectors as $\G_{J}^I =\G^A (T^A)_J^I$. 
We ignored U-parity when discussing the $\SU(\N)\to\Sp(\N)$ breaking 
in the previous section,
because U-parity was broken by $\med{\S}$.
In the present model, instead, U-parity is preserved when $\SU(\N)$ is spontaneously broken by $\med{\S}=w\mathbb{1}_{\N}$ to $\SO(\N)$.
As a result, $\SO(\N)$ confinement produces dark baryons odd under U-parity:
those built by contracting constituents  with one $\epsilon_\N$.
The lightest of such baryons is a stable DM candidate.

The $\U(1)_{\text{PQ}}$ symmetry again acts as shown in table~\ref{tab:reps}.
If $Y_\L\neq0$ all accidental global symmetries are the same as in the previous model, eq.~\eqref{u(1)simm}.
The second difference arises  if $Y_\L=0$:
leptons now acquire a Majorana mass, so U(1)$_\L$ lepton number  gets replaced by
a $\mathbb{Z}_2$ that acts as lepton parity $\L^i\to -\L^i$.
The accidental global symmetries become
$\U(1)_{\Q} \otimes \mathbb{Z}_{2} \otimes \U(1)_{\rm PQ}$.

\smallskip


\subsection{Symmetry breaking and perturbative spectrum}
In a range of potential parameters, the scalar $\S$ acquires vacuum expectation value $\med{\S}=w\mathbb{1}_{\N}$. This breaks $\SU(\N)\otimes\U(1)_{\text{PQ}}$ to $\SO(\N)$ leaving one axion and giving mass to all fermions. 
Following \cite{Buttazzo:2019mvl} the scalar field is conveniently parametrised as  
\beq
\S=\left[\left(w+\frac{s}{\sqrt{2\N}}\right)\diag(1,\dots,1)+(\Tilde{s}^b+i\Tilde{a}^b){T}_{\text{real}}^b\right]e^{\frac{ia}{\sqrt{2\N}w}}
\eeq
where $T^b_\text{real}$ are the real (symmetric) generators of $\SU(\N)$ in the fundamental representation, while $s,\tilde{s},\tilde{a},a$ are canonically normalized fields.
The mass spectrum at the perturbative level is:

\begin{itemize}
	\item $\N(\N-1)/2$ massless vectors $\mathcal{A}^a$ in the adjoint of $\SO(\N)$.
	\item $\N(\N+1)/2-1$ vectors $\mathcal{W}^b$ in the traceless symmetric of $\SO(\N)$, that acquire a squared mass $M^2_{\W}=4\g^2w^2$ eating the Goldstone bosons $\tilde{a}^b$.
	\item The massless scalar $a$, singlet under $\SO(\N)$. In view of its QCD anomalies it can be called axion and its decay constant will be $f_a=w/\sqrt{\N/2}$. 
	\item the scalar $s$, singlet under $\SO(\N)$. If symmetry breaking arises through the Coleman-Weinberg mechanism it is light with squared mass 
	$M^2_s=6(\N \lambda_\S+\lambda'_\S)w^2$.
	\item $\N(\N+1)/2-1$ scalars $\tilde{s}^b$ with squared mass $M^2_{\tilde{s}}=2(N\lambda_\S+3\lambda_\S')w^2$.
	\item One colored Dirac quark $\Psi_Q = (\Q_L,\bar{\Q}_R)^T$  with mass $M_{\Q}=y_\Q w$ in the fundamental representation of $\SO(\N)$ charged under the accidental global $\U(1)_\Q$.
	\item Three Dirac leptons $\Psi_\L^i = (\L_L^i,\bar{\L}_R^i)^T$ with masses $M_{\L^i}=y_\L^iw$ in the fundamental of $\SO(\N)$ charged under the accidental global $\U(1)_{\L_i}$ if $Y_\L\neq0$.
If $Y_\L=0$ one instead gets six Majorana leptons $\Psi_\L^i=(\L^i,\bar{\L}^i)^T$ with masses $M_{\L^i}=y_\L^iw$ in the fundamental of $\SO(\N)$ which transform as $\Psi_\L^i\to-\Psi_\L^i$ under the accidental $\mathbb{Z}_2$ symmetry.
\end{itemize}
The quarks $\Q$ are perturbatively stable thanks to the accidental global $\U(1)_{\Q}$ while the leptons $\L^i$ are stable thanks to the global $\U(1)_{\L_i}$ (if $Y_\L\neq0$) or the discrete $\mathbb{Z}_2$ symmetries (if $Y_\L=0$). 
\subsection{Confinement and bound states}
The $\SO(\N)$ gauge dynamics confines at the energy scale 
\begin{equation}
\Lambda_{\rm SO} \approx f_a\exp\left[-\frac{6\pi}{11(\N-2)\alpha_{\rm DC}(f_a)}\right].
\end{equation}
The theory contains a CP parity that extends the QCD CP parity analogously to what discussed in footnote~\ref{foot:CP}.
$\SO(\N)$ confinement leads to bound states, and we are interested in the possibly stable states.
These are the SO baryons formed by contracting constituents with an anti-symmetric $\epsilon_\N$ tensor of $\SO(\N)$,
taking into account that group theory allows SO gluons to be constituents.
\begin{itemize}
\item 
If $\N$ is even the lightest baryon is the 0-ball $\epsilon_\N \mathcal{A}^{N/2} $ made of SO gluons only, stable thanks to U-parity (see e.g.~\cite{Buttazzo:2019mvl}). 
On the other hand, the lighter baryons containing fermions 
\beq
\epsilon_{\N}\mathcal{A}^{(\N-2)/2} \Q \Q, \qquad
\epsilon_{\N}\mathcal{A}^{(\N-2)/2}\L \L, 
\qquad \epsilon_{\N}\mathcal{A}^{(\N-2)/2}\Q\L
\label{eq:baryonsNeven}
\eeq 
can decay respecting U-parity into the 0-ball plus the
corresponding lighter mesons $\Q\Q$, $\L\L$, $\Q\L$.
Such mesons are stable in the limit of exact $\U(1)_{\Q,\L}$ symmetries.

\item  If $\N$ is odd the lightest baryons contain one fermion (and thereby dubbed 1-ball)
\beq
\epsilon_{\N}\mathcal{A}^{(\N-1)/2}\Q,\qquad  \epsilon_{\N}\mathcal{A}^{(\N-1)/2}\L
\eeq
are stable if the fermion $\Q$ and/or $\L$ is stable.


\end{itemize}

\subsection{Higher dimensional operators}
Unlike in the model with an anti-symmetric discussed as in section \ref{PQpotential}, 
in the present model with a symmetric the lowest dimensional operator that breaks the PQ symmetry 
is $\det\S$ at dimension $\N$.
PQ quality now demands a weaker bound  on $f_a$,
\beq
f_a\lesssim \frac{\Lambda_{\rm UV}}{\sqrt{\N}}\left(\frac{m_\pi f_\pi}{\Lambda_{\rm UV}^2}\right)^{2/\N}\times 10^{-10/\N}.
\label{eq:faconditionSO}
\eeq
having identified $f_a= w/\sqrt{\N/2}$, since the construction of the axion 
effective Lagrangian follows exactly section \ref{sec:EFTaxion}. 
Consequently, PQ quality is now assured if $\N \gtrsim 12$ 
for both  $\Lambda_{\rm UV}\simeq M_{\rm GUT}$ and $ f_a \gtrsim 10^{9} \,\text{GeV} $ or $\Lambda_{\rm UV}\simeq M_{\rm Pl}$ and  $ f_a \gtrsim 10^{11} \,\text{GeV} $, the latter for PQ symmetry broken after inflation, otherwise $\N \gtrsim 10$ is enough.


The dimension 6 operators in eq.~\eqref{Qdecayoperators} that break fermion numbers
are allowed for $Y_\Q\pm Y_\L=\{-\frac{1}{3},\frac{2}{3},-\frac{4}{3}\}$. Assuming $M_\Q>M_\L$, the colored states $\Q$ decay with rate given in eq.~\eqref{eq: Qrate}. 
Demanding again $\tau_\Q<0.1$ sec in order to avoid problems with BBN we derive 
a bound on $f_a$ a factor of 2 stronger than in eq.\eq{fabound}.
Assuming that $\Q$ decays, the stable relics are:
\beq\label{eq: stable SO} \begin{array}{l|cc}
& \hbox{if }Y_\L \neq 0 &  \hbox{if }Y_\L = 0\\ \hline
\hbox{if $\N$ is odd} & \epsilon_\N\mathcal{A}^{(\N-1)/2}\L\hbox{ and } \L\L & \epsilon_\N\mathcal{A}^{(\N-1)/2}\L \\
\hbox{if $\N$ is even} & \epsilon_\N \mathcal{A}^{\N/2}\hbox{ and }\L\L & \epsilon_\N \mathcal{A}^{\N/2}
\end{array}.
\eeq
$Y_\L=0$ is needed to have no charged relics. 

The interplay between Landau pole, BBN and PQ-quality constraints is 
displayed in fig.\fig{tauQasymm}b, while 
for the given values of 
$Y_\Q = \{-\frac{1}{3},\frac{2}{3},-\frac{4}{3}\}$ 
and $Y_\L=0$
the predictions for the axion-photon coupling 
are the same as in fig.~\ref{fig:axionphoton}.

\subsection{$\SU(\N)/\SO(\N)$ monopoles?}
A qualitatively new feature of the $\SU(\N)\to \SO(\N)$ model is the presence of an unusual type of magnetic monopoles.
Indeed, while the model of section~\ref{Sp} had a trivial second homotopy group $\pi_2(\SU(\N)/\Sp(\N)) = 0$, 
 in the present model $\pi_2(\SU(\N)/\SO(\N)) = \mathbb{Z}_2$ is non-trivial and allows for
topologically stable $\mathbb{Z}_2$ monopoles \cite{Weinberg:1983bf}
with mass $M_{\rm mon}\approx M_\W/\alpha_{\rm DC}$.
$\mathbb{Z}_2$ monopoles differ from the well known  monopoles carrying a $\U(1)$ magnetic charge 
by the fact that their  charge is discrete, modulo $2$, so that two  $\mathbb{Z}_2$ monopole annihilate.
Their semi-classical limit was constructed in~\cite{Weinberg:1983bf,London:1985ve,Bais:1988fn}.
It was later realised that such monopoles fill multiplets under 
an emerging magnetic dual group:
in the present theory they likely fill a representation of a magnetic 
Goddard-Nuyts-Olive (GNO) \cite{Goddard:1976qe} 
dual group~\cite{hep-th/0405070,hep-th/0702102}.
Monopoles and their vectors are massless in theories where super-symmetries allow us to reliably
compute gauge dynamics beyond the semi-classical approximation~\cite{hep-th/9408099,hep-th/0702102}, but not in the present theory.

Furthermore, in our theory $\SO(\N)$ is unbroken and (most likely) confines at a lower scale $\Lambda_{\rm SO}$. 
Thereby $\SO(\N)$ magnetic fields cannot reach infinity, casting doubts on the topological  argument for monopole stability.
Indeed, it is believed that $\SO(\N)$ confinement corresponds, in the dual theory, 
to full Higgsing of the magnetic GNO dual group~\cite{hep-th/0702102},
so that monopoles can mix with electric states and decay.
We will then study cosmology assuming no stable monopoles.

\smallskip

However, as non-perturbative gauge dynamics is not firmly known,
we also consider the opposite, less likely, possibility of stable monopoles.
Then monopoles contribute to the DM density, with the abundance estimated in the rest of this section. 
Monopoles form at the $\SU(\N) \to \SO(\N)$ cosmological phase transition.
As long as $M_\W, f_a \gg  \Lambda_{\rm SO}$
the short-distance 
dynamics of monopole formation is not affected 
by $\SO(\N)$ confinement 
and the estimates for cosmological monopole production 
via the Kibble mechanism \cite{Kibble:1976sj} apply.
If the $\SU(\N) \to \SO(\N)$ phase transition is of second order, the Kibble-Zurek estimate~\cite{Zurek:1985qw,Murayama:2009nj} is 
\begin{equation}\label{eq: monoabbKZ}
\Omega_{\rm mon}h^2=1.5\times10^9\left(\frac{M_{\rm mon}}{1 \text{TeV}}\right)\left(\frac{30 T_c}{M_{\rm Pl}}\right)^\frac{3\nu}{1+\nu},
\end{equation}
where $T_c\approx M_\W$ is the critical temperature of the phase transition and $\nu$ is the related critical exponent ($\nu=1/2$ at classical level).

If instead the transition is of first order, the Kibble estimate is enhanced by a logarithmic factor due to the process of bubble nucleation~\cite{Guth:1982pn}:
\begin{equation}
\Omega_{\rm mon}h^2=1.7\times10^{11}\left(\frac{M_{\rm mon}}{1 \text{TeV}}\right)\left[\frac{T_c}{\sqrt{\sfrac{45}{(4\pi^3g_{\rm SM})}}M_\text{Pl}}\ln\left({\left({\frac{45}{4\pi^3g_{\rm SM}}}\right)^2\frac{M_\text{Pl}^4}{T_c^4}}\right)\right]^3,
\end{equation}
where $g_{\rm SM} \approx 100$ is the number of SM degrees of freedom.
Finally, the monopole abundance gets diluted by monopole annihilations and possibly 
by inflation (if PQ is broken before inflation) or by dark glue-ball decays (if PQ is broken after inflation).

\subsection{Cosmology and Dark Matter}\label{cosmoSO}
DM is composed by axions, by the 0-ball $\epsilon\mathcal{A}^{\N/2}$ (for even $\N$) or the 1-ball $\epsilon\mathcal{A}^{(\N-1)/2}\L$ (for odd $\N$ and $Y_\L=0$), and possibly by monopoles (if stable).
As in the Sp model, we consider the two possible cases.

\subsubsection{PQ broken before inflation, $T_{\rm RH}\circa{<} f_a$} 
\begin{figure}[t]
	\centering
	$$\includegraphics[width=0.40\textwidth]{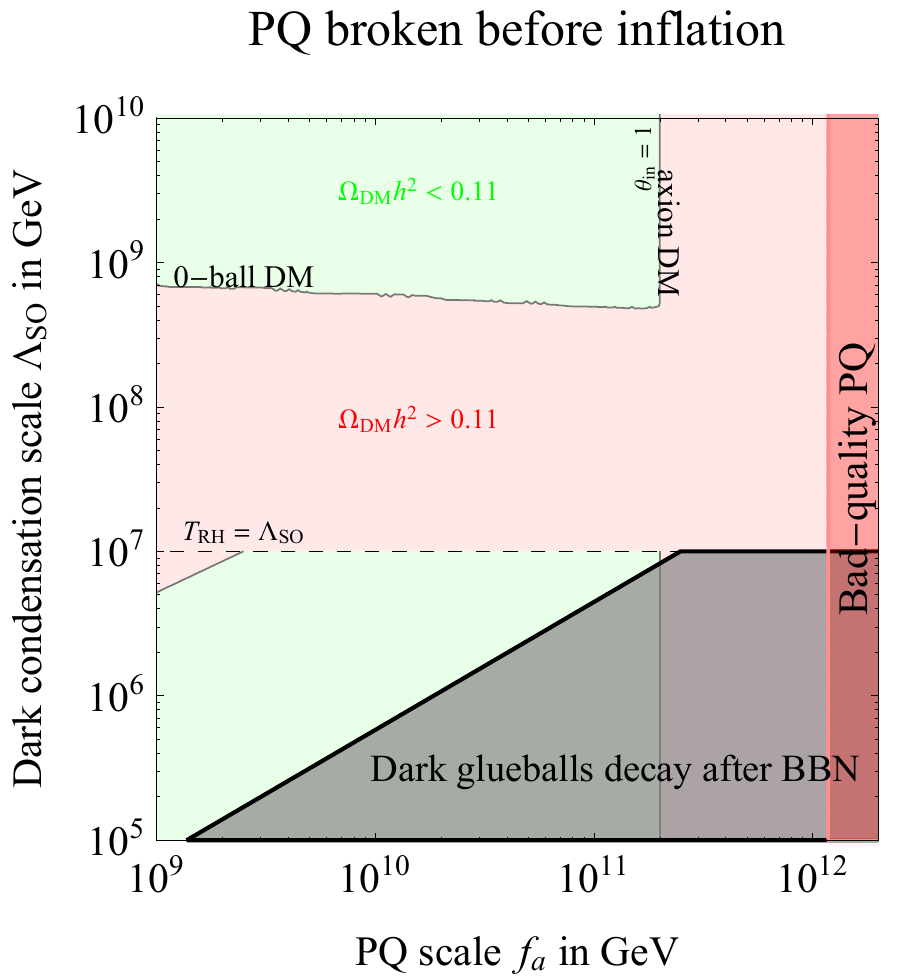}\qquad
	\includegraphics[width=0.40\textwidth]{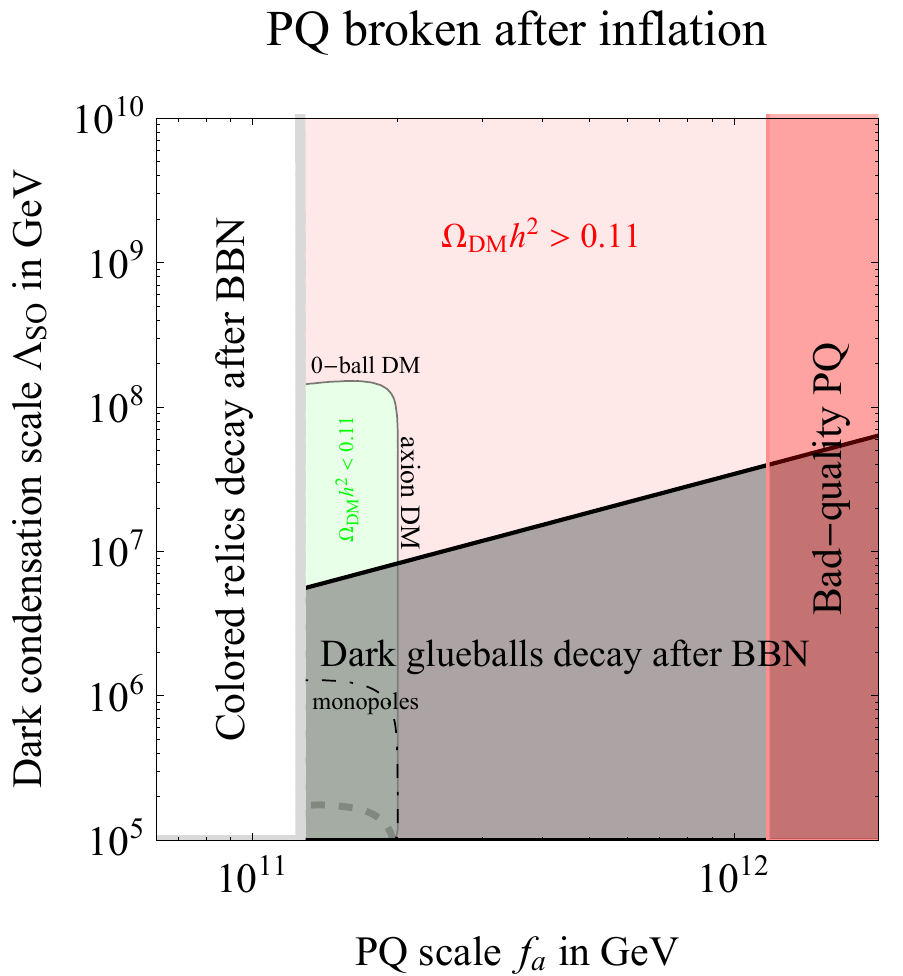}
	$$
\vspace{-1cm}
\caption{\em \label{fig:RelicPostInfSO} 
Allowed regions in the $(f_a, \Lambda_{\rm SO})$ plane for the $\SU(14)$ model with a scalar in the symmetric,
assuming $y_\Q=1$, $y_\L=0.1$, $Y_\L=0$, $Y_\Q=-1/3$.
DM is composed by axions and 0-balls. 
The relic DM abundance is reproduced along the boundaries between the red/green regions. 
White (gray) regions are excluded because $\Q$ (dark glue-balls) decay after BBN.
The region at large $f_a$ is disfavoured by poor PQ quality,  eq.~\eqref{eq:faconditionSO}. 
{\bf Left}: we assume PQ-breaking before inflation with 
$T_{\rm RH}=10^7\GeV \ll f_a$. 
The axion DM abundance is computed for $\theta_{\rm in}=1$.
{\bf Right}: PQ-breaking after inflation. 
We assumed that monopoles decay; otherwise they contribute as well 
to the DM density and the
curve along which the DM abundance is reproduced becomes the dashed curve
(Kibble-Zurek estimate) or the dot-dashed curve (Kibble-Zurek plus an estimate of monopole annihilations).}
\end{figure}

Inflation dilutes all relics, so cosmology is similar to what is discussed in section~\ref{Spbefore} for the Sp model.
Sp mesons containing $2\L$ get replaced by SO bound states containing $0\L$ or $1\L$.
Their relic abundance is again estimated assuming that such states have a non-perturbative annihilation cross section of order $1/\Lambda_{\rm SO}^2$.
Fig.\fig{RelicPostInfSO}a considers the case of even $\N=14$, showing that
there are regions where all constraints are satisfied and the cosmological abundance reproduced,
as combinations of axions and heavy relics.\footnote{If $\Lambda_{\rm SO}$ is smaller than $T_{\rm RH}$ a thermal-equilibrium population of dark gluons is present, that later dilutes the 0-ball number density, when it dominates the energy density of the Universe in the form of long-lived dark glueballs. Hence, 0-balls are under-abundant for $\Lambda_{\rm SO} \lesssim T_{\rm RH}$.}
For odd $\N$ the relics containing $1\L$ can be heavier than $\Lambda_{\rm SO}$,
and thereby have a smaller abundance produced by thermal scatterings after inflation.
Nevertheless, one can again find regions where all constraints are satisfied.


\subsubsection{PQ broken after inflation, $T_{\rm RH}\circa{>} f_a$}
Relics are now not diluted by inflation.
Still, abundances at the desired level are obtained taking into account that
they annihilate with cross section $\sigma_{\rm ann}\sim 100/\Lambda_{\rm SO}^2$
and that glue-balls can decay slowly, reheating the lighter particles and diluting the heavier relics.
We anyhow demand that $\Q$ relics decay before BBN, finding the extra bound $f_a\gtrsim  10^{11}$ GeV in fig.\fig{RelicPostInfSO}b.
Axions and  glue-balls behave as in section~\ref{Spafter}, except that now
$g_\text{DG}=\N(\N-1)$ and the scalon-glue-ball effective Lagrangian becomes
$-\sfrac{(7\N+10)\alpha_{\rm DC}(\A_{\mu\nu})^2 s}{8\pi w}$.
Fig.\fig{RelicPostInfSO}b shows  that regions exist where all constraints are satisfied, the axion quality is good and the 
cosmological DM abundance reproduced. 
We here considered even $\N=14$; similar results are found for odd $\N$.

So far we assumed that monopoles decay.  As this is not firmly established,
we also consider the possibility that stable monopoles contribute to the DM abundance.
Assuming the Kibble-Zurek estimate (dashed curve) we find that monopoles tend to be over-abundant in the allowed regions, even taking into account estimated monopole annihilations (dot-dashed curve). Considering this last case, at the boundary between the allowed and excluded regions,
the monopole abundance is 100 times larger than the DM abundance.  Unless our rough approximations over-estimate their abundance by at least this factor, monopoles are excluded.  
As order unity factor become relevant and $\N$ is somehow large,
a more reliable estimate of the monopole abundance should take into account 
that monopoles fill some representation of the GNO dual group.

\section{Conclusions}\label{concl}
We proposed two simple models that provide a high-quality accidental PQ symmetry.  
The models are based on $\SU(\N)$ gauge dynamics spontaneously broken to
either $\Sp(\N)$ or $\SO(\N)$ by a scalar $\S$ in the anti-symmetric or symmetric 
and heavy quarks and leptons in the fundamental, as summarized in table~\ref{tab:reps}.
The PQ symmetry acts as a phase rotation of $\S$ and is only broken by operators involving
the $\SU(\N)$ anti-symmetric invariant tensor with $\N$ indices.
If $\S$ is symmetric the lowest-dimensional operator that breaks the PQ symmetry
is $\det\S$ with dimension $\N$.
If $\S$ is anti-symmetric the lowest-dimensional operator that breaks the PQ symmetry
is ${\rm Pf}\,\S = \sqrt{\det\S}$ with dimension $\N/2$.
A high-quality PQ symmetry is thereby obtained for large enough $\N$.
Putting together constraints of 
theoretical type (PQ quality and no sub-Planckian Landau poles, see fig.\fig{tauQasymm})
and of
phenomenological type (colored relics decaying before BBN, cosmological DM density),
the models are viable if $\N\circa{>}12$ (for the symmetric)
or $\N\circa{>}24$ (for the anti-symmetric).

\smallskip

The models contain extra accidental symmetries that can lead to heavy relics.
Generic non-renormalizable operators break some accidental symmetries.
With an appropriate choice of heavy fermion hypercharges and masses,
heavy quarks decay before BBN into heavy leptons, that form bound states  together with dark vectors 
when the unbroken color group (SO or Sp) confines at some scale $\Lambda$.
Depending on the model, such bound states either decay or leave
cosmologically stable relics that provide extra DM candidates beyond the axion.

If the PQ symmetry is broken before inflation, extra relics get diluted away and are only 
marginally produced at reheating, so that  DM candidates can have the desired abundance.

If the PQ symmetry is broken after inflation, thermal relics of super-heavy particles are typically over-abundant.
This is not necessarily a problem, as the models under consideration contain
dark glue-balls that decay mildly slowly and can thereby partially dilute the heavier relics
(while the axion density, still in the form of vacuum energy, does not get diluted).
As a result, we find regions in the $(f_a, \Lambda)$ plane where all constraints are satisfied, as shown in
fig.\fig{RelicPreInf},\fig{RelicPostInfSO}.
Furthermore, in the $\SU(\N) \to \SO(\N)$ model, $\mathbb{Z}_2$ monopoles arise at the cosmological  phase transition.
We argued that $\SO(\N)$ confinement ruins their topological stability, but we also considered the opposite possibility of stable monopoles,
showing that they could provide extra DM through the Kibble-Zurek mechanism.


\smallskip


All in all, such models predict strongly coupled dynamics at $\Lambda \ll f_a$ and are thereby
more testable than other proposed solutions to the PQ quality problem 
(based e.g.~on discrete or abelian gauge symmetries), where the new dynamics remains confined to very high energies.
While $\Lambda$ might be larger than scales explorable at colliders,
the new physics has implications for cosmology, with possible
non-trivial cosmological interplays among different sources of DM.



Finally, we point out that the mechanism outlined in the present paper 
to address the axion quality problem could be also employed to protect the shift-symmetry 
of a generic Goldstone boson, as e.g.~in the case of ultra-light fuzzy DM~\cite{Hui:2016ltb} or cosmological approaches to the hierarchy problem (see e.g.~\cite{Strumia:2020bdy}). These scenarios typically require a shift-symmetry with a quality significantly higher than the axion case, that would be difficult to achieve otherwise, especially if the quantum-gravity breaking is not exponentially suppressed in the presence of radial modes, as claimed in~\cite{2009.03917}. The mechanism outlined here provides a natural framework to achieve such a high-quality shift-symmetry, given its robustness (the symmetry is $\N$-ality) and protecting power (the breaking can be suppressed to an arbitrarily high value by increasing $\N$). We leave a detailed study to future work.

\footnotesize

\subsubsection*{Acknowledgements}
We thank Claudio Bonati, Ken Konishi, Michele Redi and Alfredo Urbano for useful discussions.
This work was supported by the ERC grant NEO-NAT. 
The work of LDL is supported by the Marie Sk\l dowska-Curie Individual Fellowship grant AXIONRUSH (GA 840791) and the Deutsche Forschungsgemeinschaft under Germany's Excellence Strategy 
- EXC 2121 Quantum Universe - 390833306.
The work of JWW is supported by the China Scholarship Council with Grant No. 201904910660.
This research is supported in part by the MIUR under contract 2017FMJFMW (PRIN2017).

\end{document}